\documentclass[aps,pre,
longbibliography,
preprintnumbers,
twocolumn,
nofootinbib,
floatfix,
superscriptaddress,amsmath,amssymb
]{revtex4-2}
\usepackage{placeins}
\usepackage{orcidlink}
\usepackage[utf8]{inputenc}
\usepackage{graphicx}
\usepackage{amsfonts}
\usepackage{amsmath}
\usepackage{rotating}
\usepackage{amsthm}
\usepackage{autobreak}
\usepackage{placeins}
\usepackage{booktabs}
\usepackage{hyperref}
\usepackage{booktabs}
\usepackage{siunitx}
\usepackage{natbib}
\usepackage[dvipsnames]{xcolor}

\newcommand\ec{\mathit{EC}}
\newcommand{\ecs}{\ensuremath{\mathit{EC}_{\text{sym}}}}
\newcommand{\eca}{\ensuremath{\mathit{EC}_{\text{avg}}}}

\newcommand{\ecsp}{\mathrm{ec}_{\mathrm{sym}}}
\newcommand{\ecap}{\mathrm{ec}_{\mathrm{avg}}}

\newcommand{\orcidauthorLUCINI}{0000-0001-8974-8266}

\date{\today}

\begin{document}
\title{Topological signatures of criticality: the Ising case}

\author{Tak-Shing Chan\,\orcidlink{0000-0002-8302-3974}}
\email{takshingchan@gmail.com}
\affiliation{Independent Researcher}

\author{Pawe\l{} D\l{}otko\,\orcidlink{0000-0001-5352-3102}}
\email{p.dlotko@uw.edu.pl}
\affiliation{Centre of Trustworthy Artificial Intelligence for Life Sciences, University of Warsaw, Poland}

\author{Jacek Gulgowski\,\orcidlink{0000-0002-7706-9263}}
\email{j.gulgowski@uw.edu.pl}
\affiliation{Centre of Trustworthy Artificial Intelligence for Life Sciences, University of Warsaw, Poland}

\author{Biagio Lucini\,\orcidlink{\orcidauthorLUCINI}
}
\email{b.lucini@qmul.ac.uk}
\affiliation{School of Mathematical Sciences,
Queen Mary University of London,
UK}

\author{Mateusz Mas\l{}owski\,\orcidlink{0009-0009-8315-1405}}
\email{mateusz.maslowski@phdstud.ug.edu.pl}
\affiliation{University of Gdańsk and Dioscuri Centre in Topological Data Analysis, Poland}

\author{Davide Vadacchino\,\orcidlink{0000-0002-5783-5602}}
\email{davide.vadacchino@plymouth.ac.uk}
\affiliation{School of Engineering, Computing and Mathematics, University of Plymouth, UK}

\begin{abstract}
Topological properties of microscopic configurations may provide additional insight into phase transitions beyond conventional thermodynamic observables. We investigate this idea in the two-dimensional Ising model on a square and on a triangular lattice by associating cell complexes with the positive and negative spin domains and constructing an Euler-characteristic observable from their difference. We provide the expression relating this quantity to conventional observables of the Ising model. This derivation enables us to understand its expected behavior physically. As a byproduct of our calculation, we show that at any finite positive temperature the expectation of the Euler characteristic is bigger than a positive bound that does not depend on the lattice size, hence disproving a previously suggested conjecture. Monte Carlo simulations combined with multihistogram reweighting show that this spin-odd topological observable exhibits order-parameter-like behavior in the vicinity of the ferromagnetic transition. Finite-size scaling of its susceptibility yields pseudocritical temperatures that converge consistently to the known critical temperature for both lattice geometries. On the square lattice, the leading-order estimate of the critical exponent obtained from the topological observable exhibits a stronger dependence on the fitting range than the magnetization estimate. This behavior is consistent with more pronounced subleading finite-size corrections, as supported by fits including an additional correction term. On the triangular lattice, the corresponding estimate remains consistent with the theoretical Ising value. Our study suggests that topologically defined quantities such as the Euler characteristic can provide an alternative way to locate order-disorder phase transitions in the systems considered here. Since the definition of these observables is purely geometrical, they can provide useful insights in cases in which order parameters cannot be constructed.
\end{abstract}

\maketitle

\section{Introduction}

Equilibrium phase transitions are conventionally characterized by
nonanalyticities of thermodynamic functions in the thermodynamic limit and,
for continuous transitions, by the emergence of long-range correlations and
universal scaling behavior~\cite{PelissettoVicari2002}. In many systems these
changes are conveniently detected through an order parameter, such as the
magnetization in a ferromagnet. A broader question, however, is whether
structural properties of microscopic configurations can themselves provide
useful signatures of a phase transition. In particular, topology offers a natural way of characterizing changes in the structural organization of microscopic configurations while remaining insensitive to detailed metric information.

Topology has long been recognized as playing a central physical role in
phase transitions. The paradigmatic example is the
Berezinskii--Kosterlitz--Thouless transition, which is driven by the binding
and unbinding of topological defects~\cite{KosterlitzThouless1973}, but
topological properties of extended structures are informative also for
conventional order--disorder transitions. In the three-dimensional Ising
model, the approach to the critical point is accompanied by a proliferation
of handles on the interface between the two magnetized
phases~\cite{Caselle:1993yu}; in two dimensions, the Euler--Poincar\'e
characteristic of Fortuin--Kasteleyn clusters singles out the transition
point of the Potts model~\cite{BlanchardFortunatoGandolfo2002}, and the mean
Euler characteristic and mean Betti numbers of the spin domains of Ising
models with arbitrary spin display pronounced changes at the critical
point~\cite{BlanchardEtAl2006}.

A conceptually distinct line of work, usually referred to as the \emph{topological hypothesis}, relates equilibrium phase transitions to changes in the topology of energy sublevel sets in configuration space. For smooth, finite-range and confining potentials with continuous degrees of freedom, such topology changes were originally shown to be necessary for a phase transition~\cite{FranzosiPettini2004}. However, this result was challenged by a counterexample involving the two-dimensional lattice $\phi^4$ model~\cite{KastnerMehta2011}. Subsequent work has refined the theoretical framework by accounting for topological changes emerging in the thermodynamic limit~\cite{GoriEtAl2022}. The scope and limitations of this approach are also discussed in Ref.~\cite{Kastner2008}, and the framework has been extended to discrete spin systems, including the Ising model~\cite{CimasoniDelabays2019}. The advent of topological data analysis
(TDA)~\cite{Edelsbrunner2002,ZomorodianCarlsson2005,Carlsson2009,ChazalMichel2021},
and of efficient algorithms for persistent homology in
particular~\cite{Edelsbrunner,Otter2017}, turned these ideas into
quantitative computational tools, first deployed on dynamically sampled
configurations of the mean-field $XY$ and $\phi^4$
models~\cite{Donato2016}.

More recently, a novel approach has emerged in which topological invariants
are computed configuration by configuration on equilibrium ensembles, and
their expectation values are treated as statistical-mechanics observables
amenable to finite-size-scaling analysis. In condensed matter and
statistical systems, persistent-homology observables have provided
interpretable order parameters and machine-learning-free detectors of
criticality for Ising and other lattice spin
models~\cite{Cole:2020hjx,Olsthoorn:2020xzs,Tran:2021wyg}, and have yielded
quantitative estimates of critical temperatures and exponents in
two-dimensional $XY$ models~\cite{Sale2022}. In quantum field theory, the
same strategy has been used to probe center vortices and deconfinement in
$SU(2)$ gauge theory~\cite{Sale:2022qfn}, confinement in non-Abelian
lattice gauge theories~\cite{Spitz:2022tul,Spitz:2024bqh}, $\mathbb{Z}_2$
gauge theories~\cite{Sehayek:2022lxf}, dense QCD effective
models~\cite{Kashiwa:2021ctc}, and monopole networks in compact $U(1)$ and
$SU(3)$ Yang--Mills theory~\cite{Crean:2024nro,Crean:2026caz}. The present
work follows this spirit: rather than studying energy sublevel sets in the
full configuration space, we associate a cell complex with each individual
spin configuration and ask whether its topology, summarized by the Euler
characteristic of the spin domains, changes in a characteristic way across
the phase transition.

In the present work, we consider the standard two-dimensional Ising model and ask a deliberately simple question: can the topology of the positive and negative spin domains provide an alternative signature of the ferromagnetic-to-paramagnetic phase transition? To this end, for every spin configuration we associate cell complexes with the two spin domains and compute their Euler characteristics, $\ec^{+}$ and $\ec^{-}$. We then consider their half-difference, $\ecs=(\ec^{+}-\ec^{-})/2$, and their average, $\eca=(\ec^{+}+\ec^{-})/2$. The former is odd under global spin reversal and therefore provides a natural topological analog of the magnetization, whereas the latter provides a spin-reversal-even characterization of the domain topology.

We investigate these observables numerically on square and triangular lattices with periodic boundary conditions, using the magnetization as a benchmark throughout. While earlier studies investigated the Euler characteristics of individual spin domains~\cite{BlanchardEtAl2006}, our analysis focuses on their symmetrized combination as a statistical-mechanics observable. Using Monte Carlo sampling and multihistogram reweighting, we study the temperature dependence and fluctuations of the topological observables, extract pseudocritical temperatures from their susceptibility peaks, and perform a quantitative finite-size scaling analysis. This allows us to test whether the topological signal converges to the known critical point and whether its scaling behavior is robust with respect to the microscopic lattice geometry. We also derive exact relations connecting the symmetrized Euler characteristic to the magnetization and the local geometry of spin configurations, and compare its behavior in different cell-complex representations. We further show that the mean Euler characteristic density, computed using the V-construction, remains strictly positive in the thermodynamic limit at any finite, nonzero temperature, contrary to the conjecture of Blanchard et al.~\cite{BlanchardEtAl2006}. As we show below, the symmetrized Euler characteristic provides a clear and reproducible signature of the transition, although its finite-size corrections are more pronounced than those observed for the magnetization.

The remainder of the paper is organized as
follows. Section~\ref{sec:ising} introduces the Ising model and summarizes the finite-size scaling framework used in our analysis. Section~\ref{sec:topological} defines the topological
constructions and the Euler-characteristic observables associated with the positive and negative spin domains. Section~\ref{sec:numerical} describes the Monte Carlo simulations, multihistogram reweighting, and the extraction of pseudocritical points. Section~\ref{sec:results} presents the analytical relations between the observables, together with the numerical results and finite-size scaling analysis for the square and triangular lattices; in this section, we also disprove a conjecture put forward in~\cite{BlanchardEtAl2006} concerning the behavior of the Euler characteristic at low temperatures. Finally, Sec.~\ref{sec:discuss} discusses the implications and limitations of the topological observable, and
Sec.~\ref{sec:conclusions} summarizes our conclusions and perspectives for future work. 
\section{Statistical mechanics of the Ising model}
\label{sec:ising}
To fix the notation and the conventions, we provide a brief overview
of the Ising model, certainly the best known and celebrated classical spin
system. We consider the ferromagnetic Ising model on two-dimensional periodic
lattices. A spin variable $s_i\in\{-1,+1\}$ is associated with each lattice
site $i$, and the Hamiltonian is
\begin{equation}
H(\{s_i\})
=
-J\sum_{\langle ij\rangle}s_i s_j
-h\sum_i s_i ,
\label{eq:H}
\end{equation}
where $J>0$ is the nearest-neighbor ferromagnetic coupling and $h$ is the
external magnetic field. The sum $\langle ij\rangle$ runs over nearest-neighbor pairs, with four nearest neighbors per site on the square lattice and six on the triangular lattice. Throughout this work we set $k_B=J=1$ and perform
the simulations at $h=0$. For the two-dimensional square lattices considered
below, the number of spins is $N_s=L^2$.

The total magnetization is defined as
\begin{equation}
M=\sum_i s_i,
\label{eq:total_magnetization}
\end{equation}
and the magnetization per site is
\begin{equation}
m=\frac{M}{N_s}
=\frac{1}{N_s}\sum_i s_i.
\label{eq:magnetization}
\end{equation}
At inverse temperature $\beta=1/T$, the partition function is
\begin{equation}
Z(\beta,h)
=
\sum_{\{s_i\}}
e^{-\beta H(\{s_i\})},
\label{eq:partfunction}
\end{equation}
and the equilibrium probability of a configuration is
\begin{equation}
P_\beta(\{s_i\})
=
\frac{1}{Z}
e^{-\beta H(\{s_i\})}.
\label{eq:Boltzmann_ising}
\end{equation}
The free-energy density is
\begin{equation}
f(\beta,h)
=
-\frac{1}{\beta N_s}\log Z(\beta,h).
\end{equation}

At zero field, the Hamiltonian is invariant under the global spin reversal
$s_i\mapsto -s_i$. In the thermodynamic limit this $\mathbb{Z}_2$ symmetry
is spontaneously broken below the critical temperature $T_c$, with the
magnetization serving as the conventional order parameter. For a finite
system at $h=0$, however, spin-reversal symmetry implies
$\langle M\rangle=0$. We therefore use $\langle |M|\rangle$ when displaying
the finite-volume order parameter.

The magnetic susceptibility per site is related to magnetization fluctuations by
\begin{equation}
    \begin{aligned}
        \chi_m
        &=\left.\frac{\partial \langle m\rangle}{\partial h}\right|_{h=0}
        =\beta N_s\left(\langle m^2\rangle-\langle m\rangle^2\right)
        \\
        &=\frac{\beta}{N_s}\left(\langle M^2\rangle-\langle M\rangle^2\right).
    \end{aligned}
\label{eq:magnetic_susceptibility}
\end{equation}

More generally, for an intensive observable $O$ we define the corresponding fluctuation susceptibility as
\begin{equation}
\chi_O=\beta N_s\left(\langle O^2\rangle-\langle O\rangle^2\right).
\label{eq:observable_susceptibility}
\end{equation}
This quantity will be used below for both conventional and topological observables.

For finite systems at zero field, where the two magnetized sectors are sampled symmetrically, it is often convenient in numerical finite-size analyses to use the finite-volume fluctuation estimator
\begin{equation}
\chi_m^{(L)}=\beta N_s\left(\langle m^2\rangle-\langle |m|\rangle^2\right).
\label{eq:magnetic_susceptibility_finite}
\end{equation}

The corresponding estimators used for the observables considered in the finite-size analysis are specified in Sec.~\ref{sec:numerical}.

The two-dimensional ferromagnetic Ising model undergoes a continuous phase
transition. For the square lattice the exact critical temperature is
\begin{equation}
T_c^{\mathrm{sq}}
=
\frac{2}{\log(1+\sqrt{2})}
\simeq 2.269185,
\label{eq:Tc_square}
\end{equation}
while for the triangular lattice it is
\begin{equation}
T_c^{\mathrm{tri}}
=
\frac{4}{\log 3}
\simeq 3.640957
\label{eq:Tc_triangular}
\end{equation}
in the units used here~\cite{Onsager1944,Houtappel1950}.
Both models belong to the two-dimensional Ising universality class. Close to the critical point, the correlation length diverges as
\begin{equation}
\xi \sim |t|^{-\nu},
\qquad
t=\frac{T-T_c}{T_c},
\label{eq:correlation_length_scaling}
\end{equation}
which defines the correlation-length critical exponent $\nu$. For the two-dimensional Ising universality class,
\begin{equation}
\nu=1.
\label{eq:2d_ising_exponent}
\end{equation}
This value is exact and independent of the lattice geometry within the same universality class~\cite{PelissettoVicari2002}.

\subsection{Finite-size scaling}
\label{sec:crit_scaling}

In a finite system of linear size $L$, the divergence of the correlation length is cut off when $\xi$ becomes comparable with $L$. As a consequence, thermodynamic singularities are rounded and their locations become size dependent. Finite-size scaling relates this dependence to the critical exponents~\cite{FisherBarber1972}.

For a finite-size observable $O$, let $T_O^c(L)$ denote the pseudocritical temperature identified from the maximum of its fluctuation susceptibility. To leading order, its shift from the thermodynamic critical
temperature scales as
\begin{equation}
T_O^c(L)-T_c \sim L^{-1/\nu}.
\label{eq:fit_Tc_shift_prop}
\end{equation}
Subleading corrections generally produce deviations from these asymptotic
forms at smaller lattice sizes.

In the numerical analysis below, the magnetization provides a benchmark
for this finite-size scaling procedure. We then test whether the
pseudocritical temperatures defined from the symmetrized Euler
characteristic, $\ecs$, follow the same leading shift form. Importantly,
this scaling behavior is not assumed to hold exactly for the topological
observable at finite $L$; its sensitivity to corrections to scaling is one
of the questions investigated in this work.
\section{Topological observables}
\label{sec:topological}
In this section we introduce some general concepts in homology theory and 
then specialize them to the case of configurations of the Ising model.
\subsection{Cubical complexes, simplicial complexes, Euler characteristics}

To work with homology theory in the context of the Ising model defined on a square lattice, it is natural to provide a basic exposition
of the properties of \emph{cubical
complexes}. The reader will find a more detailed and formal introduction to cubical complexes and their homology in~\cite{kaczynski2004}. 

We will denote with $[n,n+1]$ for 
$n \in \mathbb{N}$ as a \emph{non-degenerated elementary interval} and
with $[n,n]$ a \emph{degenerated
elementary interval}. An \emph{elementary cube} is then simply a Cartesian
product of (degenerated or not) elementary intervals. 
The number of non-degenerated intervals in the product is a \emph{dimension of a cube}, 
while the total number of the elementary intervals in the product is 
the \emph{ambient dimension} of the
cube. 

An important object which relates a cube of dimension $n > 0$ to its 
boundary is the \emph{boundary operator}. In this paper, for simplicity,
we will consider the boundary operator only over the $\mathbb{Z}_2$ field. Given a non-degenerated elementary interval $[n,n+1]$, its boundary is 
defined as $\partial [n,n+1]= [n+1,n+1]+[n,n]$. For a degenerated elementary
interval $[n,n]$ we set $\partial [n,n] = 0$. 

Given a cube
$C = I_1 \times I_2 \times \ldots \times I_n$ its boundary acts on the
factors in turn, 
$\sum_{i=1}^n I_1 \times I_2 \times \ldots \times \partial I_i \times \ldots
\times I_n$, and results in a formal sum of cubes. Such a formal sum of cubes
with $\mathbb{Z}_2$ coefficients is an instance of a type of object called
\emph{chain}. 
It is
a standard to check that $\partial \partial = 0$. 

In addition to cubical complexes that are well-suited to hypercubic lattices, we will also make use of a second class of combinatorial objects suited for lattices with a different microscopic structure. These are \emph{simplicial complexes}. Again, we will only provide an elementary introduction. For a more detailed and formal treatment, please consult Ref~\cite{Edelsbrunner}.

A \emph{$k$--simplex} will be defined as the convex hull of $k+1$ affinely independent points $v_0$, $v_1$, ..., $v_k$ and will be denoted by $[v_0, v_1, \ldots, v_k]$.
Its faces are obtained by taking the convex hulls of all nonempty subsets of its vertex set. 
Hence vertices, edges, and filled triangles correspond to simplices of dimension $0$, $1$, and $2$, respectively.

Working over the field $\mathbb{Z}_2$, the boundary of a simplex
$S = [v_0, v_1, \ldots, v_k]$ is defined as the sum of all $(k-1)$--simplices obtained by removing one vertex:
\begin{align}
\begin{autobreak}
\partial S =
[v_1, v_2, \ldots, v_k]
+ [v_0, v_2, \ldots, v_k]
+ \cdots
+ [v_0, v_1, \ldots, v_{k-1}].
\end{autobreak}
\end{align}
Extending this rule linearly gives the boundary operator on chains, and, as in the cubical case, one verifies that $\partial^2 = 0$.

A \emph{simplicial complex} $\mathcal{K}$ is a collection of simplices satisfying two conditions:
\begin{enumerate}
    \item if $S \in \mathcal{K}$, then every face $F$ of $S$ also belongs to $\mathcal{K}$;
    \item the intersection of any two simplices in $\mathcal{K}$ is either empty or a face of each.
\end{enumerate}
The dimension of $\mathcal{K}$ is the maximal dimension of a simplex contained in it. 

Although their construction differs from the cubical setting introduced earlier,
simplicial complexes serve the same purpose: they provide a finite combinatorial
toolbox suitable for computing homological invariants.
\subsection{Homology}

Given a cubical or simplicial complex $\mathcal{K}$, let $\mathcal{K}_i$ denote
the set of all $i$--dimensional cells in $\mathcal{K}$.  
The $i$th chain group $C_i(\mathcal{K})$ consists of all $\mathbb{Z}_2$–linear
combinations of elements of $\mathcal{K}_i$.  
The boundary operator extends linearly to chains: if
\begin{equation}
c = \sum_{S \in \mathcal{K}_i} \alpha_S\, S,
\end{equation}
then
\begin{equation}
\partial c = \sum_{S \in \mathcal{K}_i} \alpha_S\, \partial S.
\end{equation}
Thus, $\partial : C_i(\mathcal{K}) \rightarrow C_{i-1}(\mathcal{K})$ for every $i>0$, and in all cases $\partial^2 = 0$.

An $i$--dimensional chain $c$ is called a \emph{cycle} if $\partial c = 0$, and
a \emph{boundary} if there exists $d \in C_{i+1}(\mathcal{K})$ with $\partial d
= c$.  Every boundary is automatically a cycle, again because $\partial^2 = 0$.  
The subgroups of cycles and boundaries are denoted
\begin{equation}
Z_i(\mathcal{K}) = \ker \partial, \qquad
B_i(\mathcal{K}) = \operatorname{im} \partial.
\end{equation}

Homology theory classifies cycles up to boundaries.  
The $i$th homology group is defined as
\[
H_i(\mathcal{K}) = Z_i(\mathcal{K}) / B_i(\mathcal{K}),
\]
and intuitively measures $i$--dimensional “holes’’ in the complex:  
connected components for $i=0$, tunnels for $i=1$, cavities for $i=2$, and so on.  
Two cycles are \emph{homologous} if they differ by a boundary.  

The rank of $H_i(\mathcal{K})$, denoted $\beta_i(\mathcal{K})$, is the $i$th \emph{Betti number}.

A somewhat more fundamental, but very informative, way of encapsulating
information about Betti numbers is an \emph{Euler characteristic},
$\ec(\mathcal{K}) = \sum_{i=0}^{dim(\mathcal{K})} (-1)^i \beta_i$. 
At the same time $\ec(\mathcal{K})$ is equal to the alternating sum of number of
cells of dimension zero minus number of cells of dimension one plus number of
cells in dimension two etc, formally $\ec(\mathcal{K}) =
\sum_{i=0}^{dim(\mathcal{K})} (-1)^i | \mathcal{K}_i |$, where $|\cdot|$ denotes
the cardinality of the complex. The later formula give a clear and efficient
way of computing the Euler characteristic from a cubical or simplicial complex
alone without a need for Homology groups computations. 
\subsection{From Ising model to cubical complexes}
There are two natural ways of associating a cubical complex with a configuration of the Ising model. They differ in the geometric object assigned to an individual spin. We will refer to them as the \emph{vertex construction} (\emph{V-construction}) and the \emph{top-cell construction} (\emph{T-construction}).

In the V-construction, each lattice site, and hence each spin of the Ising configuration, is identified with a vertex of the underlying cubical lattice. In the T-construction, each spin is instead associated with a top-dimensional cubical cell. Thus, in two dimensions a spin corresponds respectively to either a vertex or an elementary square, while in $d$ dimensions the corresponding top-dimensional object in the T-construction is a $d$-dimensional cube.

In both constructions one must additionally choose which of the two spin values is used to generate the complex. Since the Ising spins take values $+$ and $-$, either spin domain may be selected. We do not assign any a priori physical preference to either choice and therefore construct both the positive and negative complexes.

Throughout these constructions we impose periodic boundary conditions. The underlying cubical lattice is therefore regarded as a cubical decomposition of the $d$-dimensional torus $T^d$, which may be represented by the fundamental domain
\[
[l_1^{\min},l_1^{\max}]
\times
[l_2^{\min},l_2^{\max}]
\times \cdots \times
[l_d^{\min},l_d^{\max}],
\]
with opposite faces identified.

\subsubsection{V-construction}
Each lattice site, and hence each spin of the Ising configuration, corresponds to a vertex of the cubical lattice located at the same position. For a chosen spin value, we regard the corresponding vertices as active. For example, choosing the positive spins gives a cubical complex $\mathcal{K}$ whose vertices are those associated with spins of value $+$, together with all edges and higher-dimensional cubes whose vertices are all active.

The \emph{positive complex}, $\mathcal{K}^{+}$, is obtained by taking the vertices corresponding to positive spins as active and including all cubical cells whose vertices are active. Analogously, the \emph{negative complex}, $\mathcal{K}^{-}$, is obtained by taking the vertices corresponding to negative spins as active. In this way, the two complexes describe the geometric structure of the two spin domains on equal footing.

Let $\mathcal{L}$ denote the cubical complex associated with the full underlying lattice. The positive and negative complexes are complementary at the level of their vertices: every vertex of the underlying cubical lattice belongs to exactly one of the two complexes. Thus,
\[
V(\mathcal{K}^{+}) \sqcup V(\mathcal{K}^{-}) = V(\mathcal{L}).
\]
This complementarity does not, however, extend to higher-dimensional cells. An edge, or more generally a cubical cell, whose vertices contain both positive and negative spins belongs to neither $\mathcal{K}^{+}$ nor $\mathcal{K}^{-}$. Consequently, the union of the two complexes does not in general reproduce the full cubical decomposition of the lattice.

\begin{figure}
\centering
\includegraphics[width=\columnwidth]{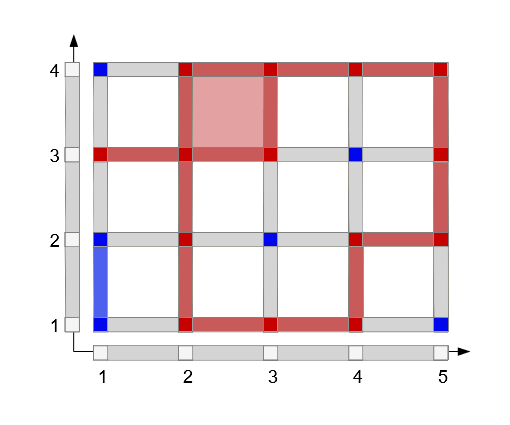}
    \caption{A simple example illustrating that the number of holes in one cubical complex does not uniquely determine the number of connected components of the opposite complex enclosed within them. The positive (red) complex contains a nontrivial one-dimensional cycle enclosing two distinct connected components of the negative (blue) complex, supported at the vertices $[3]\times[2]$ and $[4]\times[3]$. Since a single cycle can also enclose just one connected component, there is no one-to-one correspondence between these two counts.}
    \label{lack_of_duality}
\end{figure}

The geometry of the spin domains suggests a possible relation between $\beta_1$ of one complex and $\beta_0$ of the other: a hole in the positive complex contains spins of the opposite sign, and one might expect it to correspond to a connected component of the negative complex. In the square-lattice V-construction, however, the negative spins within a single hole need not form a connected complex. As illustrated in Figure~\ref{lack_of_duality}, one hole in the positive complex can contain several disconnected components of the negative complex. The same observation applies with the spin signs exchanged. Thus, there is no one-to-one correspondence between holes in one complex and connected components of the other. Moreover, on the torus, $\beta_1$ also counts independent cycles winding around the periodic directions, which do not necessarily enclose a region.

\subsubsection{T-construction}

We now consider the T-construction. In this case, each lattice site, and hence each spin of the Ising configuration, is associated with a top-dimensional cubical cell. In $d$ dimensions this is a $d$-dimensional cube, while in two dimensions each spin is represented by an elementary square.

For a chosen spin value, the corresponding top-dimensional cells are included in the complex together with all of their boundary cells. Thus, the positive complex $\mathcal{K}^{+}$ is obtained by including the top-dimensional cells associated with positive spins, together with all their faces, while the negative complex $\mathcal{K}^{-}$ is constructed analogously from the negative spins. Including the complete boundaries of the selected top-dimensional cells ensures that the resulting objects are cubical complexes.

In contrast to the V-construction, the positive and negative T-complexes are complementary at the level of their top-dimensional cells: every top-dimensional cell belongs to exactly one of the two complexes. Consequently, their union contains all top-dimensional cells of the underlying cubical decomposition. Since each such cell is included together with its boundary, the two complexes together cover the full cubical complex,
\[
\mathcal{K}^{+}\cup\mathcal{K}^{-}=\mathcal{L}.
\]
The complexes themselves are not, however, disjoint. Top-dimensional cells associated with opposite spin values may share lower-dimensional boundary cells, and therefore
\[
\mathcal{K}^{+}\cap\mathcal{K}^{-}\neq\varnothing
\]
in general.

As in the V-construction, one may expect a relation between the Betti numbers of the positive and negative complexes. In two dimensions, a nontrivial one-dimensional cycle in one complex may surround a region occupied by top-dimensional cells of the opposite complex, suggesting a possible correspondence between $\beta_1$ of one complex and $\beta_0$ of the other.

\begin{figure}
\centering
\includegraphics[scale=0.5]{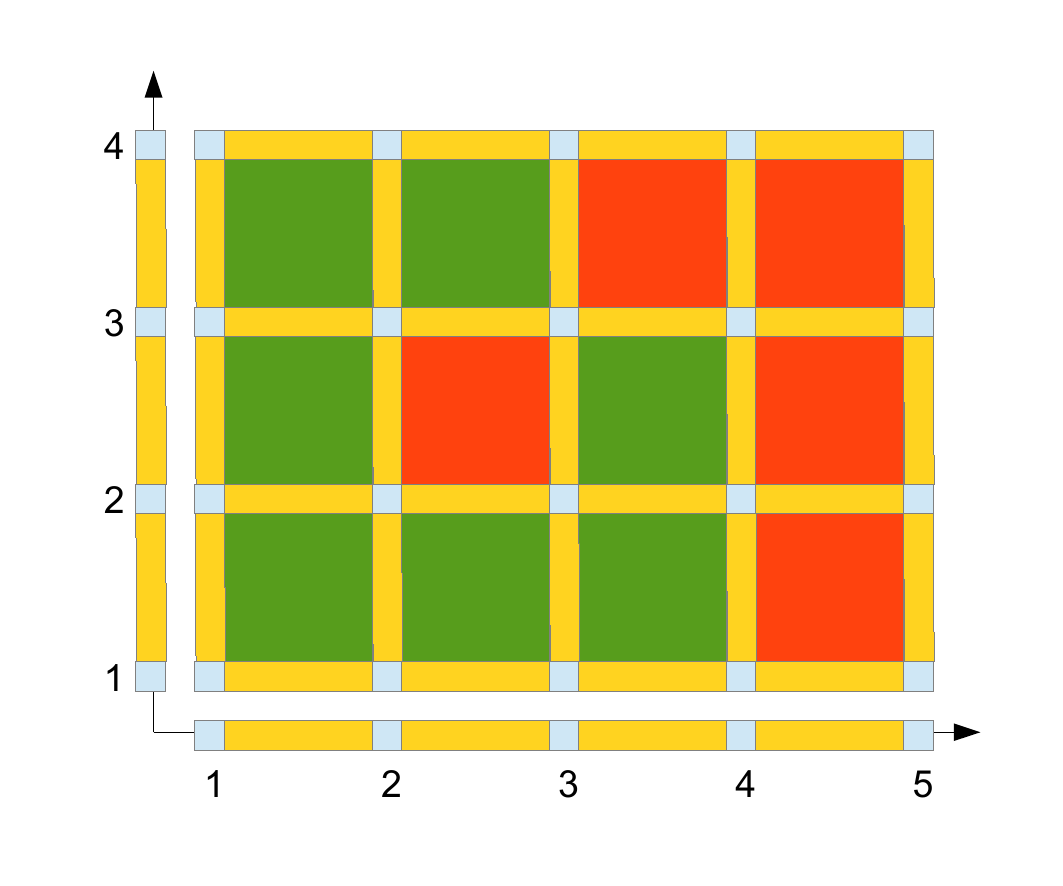}
    \caption{A simple example illustrating that a hole in one complex need not correspond to a separate connected component of the opposite complex in the T-construction. Suppose that the cubes marked in green belong to the positive complex and those marked in red to the negative complex. The positive complex contains a nontrivial one-dimensional cycle surrounding the negative cube $[2,3]\times[2,3]$. However, this cube is connected to the remaining part of the negative complex, through the common vertex $[3]\times[3]$. Thus, although the two regions meet only at a corner, they belong to the same connected component of the negative cubical complex. The enclosed negative cube therefore does not contribute a separate connected component to $\beta_0$ of the negative complex.}
    \label{lack_of_duality2}
\end{figure}

Such a correspondence is again not one-to-one. The inclusion of all boundary cells allows top-dimensional cells of the same complex to become connected through lower-dimensional cells. In particular, two regions may meet only at a single vertex and nevertheless belong to the same connected component of the cubical complex. As a result, a region enclosed by a nontrivial cycle of one complex need not constitute a separate connected component of the opposite complex. A simple example is shown in Figure~\ref{lack_of_duality2}.

Therefore, although the positive and negative T-complexes are complementary at the level of their top-dimensional cells, their Betti numbers do not satisfy a simple duality relation.

\subsection{From Ising model to simplicial complexes}
In addition to cubical complexes arising from the square lattice, we also consider the two-dimensional Ising model defined on a triangular lattice. In this case, the natural topological representation is a simplicial complex. As in the square-lattice construction, each spin is associated with a vertex located at the corresponding lattice site. For a chosen spin value, the vertices corresponding to spins of that sign are regarded as active. Unlike in the cubical case, we restrict attention to this vertex-based construction and use it throughout the simplicial analysis.

Two active vertices are connected by an edge whenever the corresponding lattice sites are nearest neighbors. Whenever the three vertices of an elementary triangle of the lattice are active, the corresponding two-dimensional simplex is included as well. This defines a simplicial subcomplex of the underlying triangular lattice.

As in the cubical case, we do not assign a preferred role to either spin value. We therefore construct two simplicial complexes for each Ising configuration: the positive complex, obtained by taking the vertices corresponding to positive spins as active, and the negative complex, obtained analogously from the negative spins.

In contrast to the cubical constructions discussed above, the simplicial structure of the triangular lattice avoids the local corner-type connections that obstruct a direct relation \cite{Hatcher2002} between the topology of the two spin domains. Let $K^+$ and $K^-$ denote the positive and negative spin complexes. When both complexes are nonempty, Alexander--Poincar\'e duality implies
\begin{equation}
\beta_0(K^-)-1=\beta_1(K^+)-r_+, \qquad \beta_0(K^+)-1=\beta_1(K^-)-r_-,
\label{eq:triangular_duality}
\end{equation}
where $r_\pm\in\{0,1,2\}$ is the number of independent homology classes of cycles in $K^\pm$ that remain nontrivial in the ambient torus $T^2$. These terms account for cycles winding around the periodic directions, which need not separate distinct components of the opposite spin complex. For the complementary spin domains on the triangular lattice, these ranks satisfy $r_++r_-=2$. Together with Eq.~\ref{eq:triangular_duality} and $\beta_2(K^\pm)=0$ when both complexes are nonempty, this gives $\ec^+=-\ec^-$. If one complex is empty, the other is the full torus, and both Euler characteristics are again zero.

\subsection{Euler observables}

For each Ising configuration, we evaluate separately the Euler characteristics of the positive and negative complexes, denoted by $\ec^{+}$ and $\ec^{-}$, respectively. Since the choice of which spin value is labeled positive or negative is a matter of convention, it is natural to combine these quantities into combinations that transform simply under the exchange of the two spin domains.

The symmetrized Euler characteristic is defined as
\begin{equation}
\ecs = \frac{\ec^{+}-\ec^{-}}{2}.
\end{equation}
Under the exchange of positive and negative spins, $\ecs$ changes sign.

The averaged Euler characteristic is defined as
\begin{equation}
\eca = \frac{\ec^{+}+\ec^{-}}{2}.
\end{equation}
In contrast, $\eca$ is invariant under the exchange of the two spin domains.

Since both Euler characteristics are extensive quantities, we also define their densities per lattice site,
\begin{equation}
\ecsp=\frac{\ecs}{N_s},
\qquad
\ecap=\frac{\eca}{N_s}.
\label{eq:euler_densities}
\end{equation}
These intensive quantities are used in the numerical analysis and in the definition of the corresponding finite-volume fluctuation estimators. For the square lattice, twice the averaged Euler characteristic density in the V-construction, $2\ecap^V$, corresponds to the global Euler characteristic density considered by Blanchard et al.~\cite{BlanchardEtAl2006} for the standard Ising model ($Q=1$).

For the simplicial construction on the triangular lattice, as established above, the positive and negative complexes satisfy
\begin{equation}
\ec^{+}=-\ec^{-},
\end{equation}
as a consequence of the duality between the two spin domains on the ambient torus. It follows immediately that
\begin{equation}
\eca=0,
\end{equation}
for every spin configuration, while
\begin{equation}
\ecs=\ec^{+}=-\ec^{-}.
\end{equation}
Consequently, when the absolute value of the symmetrized Euler characteristic is considered,
\begin{equation}
|\ecs|=|\ec^{+}|=|\ec^{-}|.
\end{equation}
Thus, for the triangular lattice, $\eca$ is identically zero and contains no additional information, whereas $\ecs$ fully determines the Euler characteristics of both spin complexes.

We now return to the cubical complexes associated with the square lattice. Although the V- and T-constructions generally assign different cubical complexes to the same Ising configuration, we show that the symmetrized Euler characteristic $\ecs$ is identical in the two constructions.

Let
\[
F_k^{V,\pm}
\]
and
\[
F_k^{T,\pm}
\]
denote the numbers of $k$-dimensional cells in the positive and negative complexes obtained using the V- and T-constructions, respectively. We define
\[
\Delta F_k^V
=
F_k^{V,+}-F_k^{V,-},
\]
and
\[
\Delta F_k^T
=
F_k^{T,+}-F_k^{T,-}.
\]

For either construction, the symmetrized Euler characteristic is defined as
\[
\ecs
=
\frac{\ec_+-\ec_-}{2}.
\]
Since in two dimensions
\[
\ec_\pm
=
F_0^\pm-F_1^\pm+F_2^\pm,
\]
we obtain
\[
2\ecs
=
\left(F_0^+-F_0^-\right)
-
\left(F_1^+-F_1^-\right)
+
\left(F_2^+-F_2^-\right),
\]
or equivalently,
\[
2\ecs
=
\Delta F_0-\Delta F_1+\Delta F_2.
\]

We now compare the three differences $\Delta F_k$ in the V- and T-constructions.

First, recall the local geometric correspondence between the two constructions. In the V-construction, each spin is represented by a vertex. An edge is determined by a pair of neighbouring spins, while a two-dimensional square is determined by four spins at its vertices. In the T-construction, each spin is instead represented by a two-dimensional square. An edge lies between two neighbouring squares, while a vertex is incident to four squares. Thus, the roles of zero- and two-dimensional cells are exchanged between the two constructions.

We first compare the vertices of the V-construction with the two-dimensional cells of the T-construction. Both are in one-to-one correspondence with individual spins. A positive spin gives one positive vertex in the V-construction and one positive square in the T-construction, while a negative spin gives one negative vertex and one negative square. Hence
\[
F_0^{V,+}
=
F_2^{T,+},
\]
and
\[
F_0^{V,-}
=
F_2^{T,-}.
\]
Subtracting these two relations gives
\[
F_0^{V,+}-F_0^{V,-}
=
F_2^{T,+}-F_2^{T,-},
\]
and therefore
\[
\Delta F_0^V
=
\Delta F_2^T.
\]

We next compare the edge contributions. Let
\[
E_{++},
\qquad
E_{--},
\qquad
E_{+-}
\]
denote the numbers of nearest-neighbour pairs with spins $(++)$, $(--)$, and mixed sign, respectively.

In the V-construction, an edge is present in the positive complex only when both of its endpoints carry positive spins, and in the negative complex only when both endpoints carry negative spins. Therefore,
\[
F_1^{V,+}=E_{++},
\]
and
\[
F_1^{V,-}=E_{--}.
\]
Hence,
\[
\Delta F_1^V
=
F_1^{V,+}-F_1^{V,-}
=
E_{++}-E_{--}.
\]

In the T-construction, an edge belongs to the positive complex whenever at least one of the two adjacent top-dimensional cells corresponds to a positive spin. Thus,
\[
F_1^{T,+}
=
E_{++}+E_{+-}.
\]
Analogously,
\[
F_1^{T,-}
=
E_{--}+E_{+-}.
\]
The mixed-spin edges are therefore contained in both T-complexes and cancel when the difference between the positive and negative edge counts is taken:
\[
\begin{aligned}
\Delta F_1^T
&=
F_1^{T,+}-F_1^{T,-}
\\
&=
\left(E_{++}+E_{+-}\right)
-
\left(E_{--}+E_{+-}\right)
\\
&=
E_{++}-E_{--}.
\end{aligned}
\]
Comparing with the V-construction gives
\[
\Delta F_1^T
=
\Delta F_1^V.
\]

Finally, we compare the vertices of the T-construction with the two-dimensional cells of the V-construction. A vertex in the T-construction is incident to four top-dimensional cells and is therefore associated with the same four spins that determine a square in the V-construction.

Let
\[
N_+,
\qquad
N_-,
\qquad
N_m
\]
denote the numbers of local $2\times 2$ spin configurations which are
respectively, entirely positive, entirely negative, or mixed.

In the V-construction, a two-dimensional cell belongs to the positive
complex only when all four of its vertices carry positive spins, and to
the negative complex only when all four carry negative spins. Therefore,
\[
F_2^{V,+}=N_+,
\]
and
\[
F_2^{V,-}=N_-.
\]
Hence,
\[
\Delta F_2^V
=
N_+-N_-.
\]

In the T-construction, a vertex belongs to the positive complex whenever
at least one of the four incident top-dimensional cells corresponds to a
positive spin, and analogously for the negative complex. Therefore,
\[
F_0^{T,+}
=
N_+ + N_m,
\]
while
\[
F_0^{T,-}
=
N_- + N_m.
\]
The vertices corresponding to mixed local spin configurations are thus
contained in both T-complexes and cancel in the difference. Consequently,
\[
\begin{aligned}
\Delta F_0^T
&=
\left(N_+ + N_m\right)
-
\left(N_- + N_m\right)
\\
&=
N_+ - N_-.
\end{aligned}
\]

Comparing the two expressions gives
\[
\Delta F_0^T
=
\Delta F_2^V.
\]

Collecting the three relations, we obtain
\[
\Delta F_0^V=\Delta F_2^T,
\qquad
\Delta F_1^V=\Delta F_1^T,
\qquad
\Delta F_2^V=\Delta F_0^T.
\]

The symmetrized Euler characteristic in the V-construction is
\[
2\ecs^V
=
\Delta F_0^V
-
\Delta F_1^V
+
\Delta F_2^V,
\]
while in the T-construction it is
\[
2\ecs^T
=
\Delta F_0^T
-
\Delta F_1^T
+
\Delta F_2^T.
\]

Using the relations derived above,
\[
2\ecs^T
=
\Delta F_2^V
-
\Delta F_1^V
+
\Delta F_0^V.
\]
Reordering the terms gives
\[
2\ecs^T
=
\Delta F_0^V
-
\Delta F_1^V
+
\Delta F_2^V
=
2\ecs^V.
\]
Therefore,
\[
\ecs^T
=
\ecs^V.
\]

Thus, the symmetrized Euler characteristic does not depend on whether the spin configuration is represented using the V- or T-construction. In this sense, $\ecs$ defines an observable of the spin configuration itself rather than of the particular cubical realization used to represent it.

We can similarly determine how the averaged Euler characteristic transforms
under the change from the V- to the T-construction. Since
\[
2\eca
=
\ec^+ + \ec^-,
\]
we have
\[
2\eca^T
=
F_0^{T,+}+F_0^{T,-}
-
\left(F_1^{T,+}+F_1^{T,-}\right)
+
F_2^{T,+}+F_2^{T,-}.
\]

Using the notation introduced above, the numbers of vertices in the two
T-complexes are
\[
F_0^{T,+}=L^2-N_-,
\qquad
F_0^{T,-}=L^2-N_+.
\]
Indeed, a vertex is absent from the positive T-complex only when all four
incident top-dimensional cells correspond to negative spins, and
analogously for the negative complex.

For the edges,
\[
F_1^{T,+}=E_{++}+E_{+-},
\qquad
F_1^{T,-}=E_{--}+E_{+-},
\]
while every spin corresponds to exactly one top-dimensional cell, so that
\[
F_2^{T,+}+F_2^{T,-}=L^2.
\]
Since the square lattice with periodic boundary conditions contains
$2L^2$ edges,
\[
E_{++}+E_{--}+E_{+-}=2L^2.
\]
It follows that
\[
F_1^{T,+}+F_1^{T,-}
=
2L^2+E_{+-}.
\]
Therefore,
\begin{equation}
        \begin{aligned}
            2\eca^T
            &=
            2L^2-(N_++N_-)
            -
            (2L^2+E_{+-})
            +
            L^2
            \\
            &=
            L^2-(N_++N_-)-E_{+-}.
        \end{aligned}
    \label{eq:eca_T}
\end{equation}

For the V-construction,
\[
F_0^{V,+}+F_0^{V,-}=L^2,
\]
\[
F_1^{V,+}+F_1^{V,-}
=
E_{++}+E_{--}
=
2L^2-E_{+-},
\]
and
\[
F_2^{V,+}+F_2^{V,-}=N_++N_-.
\]
Hence,
\[
\begin{aligned}
2\eca^V
&=
L^2
-
(2L^2-E_{+-})
+
(N_++N_-)
\\
&=
-L^2+E_{+-}+N_++N_-.
\end{aligned}
\]
Comparing the two expressions gives
\begin{equation}
    \eca^T=-\eca^V.
    \label{ec_avg:odd}
\end{equation}

Since the choice between the V- and T-constructions leaves $\ecs$ unchanged and affects $\eca$ only by an overall sign, the two constructions carry the same information for the Euler observables considered here. We therefore restrict the following analysis to the T-construction.
\section{Numerical Methods}\label{sec:numerical}

\subsection{Update algorithm}
The properties of the Ising model can be explored numerically using Markov chain Monte Carlo methods. Successive configurations generated by a Markov chain are generally correlated, and these correlations become particularly important near a critical point because of critical slowing down.

To reduce autocorrelations near the transition, we use the Wolff cluster algorithm~\cite{wolff1989}, combined with Metropolis-Hastings~\cite{LandauBinder} updates. In a Wolff update, a seed spin is chosen at random and a cluster of aligned nearest-neighbor spins is grown by activating bonds with probability
\begin{equation}
p = 1 - e^{-2\beta J}.
\end{equation}
All spins belonging to the resulting cluster are then flipped simultaneously. This non-local update substantially reduces critical slowing down compared with purely local updates.

In the simulations used here, one hybrid update step consists of one complete Metropolis-Hastings sweep over the lattice followed by 15 Wolff cluster updates. After thermalization, observables are measured from the recorded configurations according to
\begin{equation}
\bar{O}
=
\frac{1}{N}
\sum_{t=1}^{N}
O(\{s_i\}_t),
\label{eq:MCavg}
\end{equation}
where $\{s_i\}_t$ denotes the $t$th recorded configuration and $N$ is the number of recorded configurations.
\subsection{Collecting observables}
For each generated Ising configuration, we construct the positive and negative complexes using the constructions introduced in Sec.~\ref{sec:topological} and evaluate their topological properties. In the cubical case, both the V- and T-constructions are available. Since $\ecs$ is the primary Euler observable considered in our analysis and, as shown above, is independent of the choice between these two constructions, we use the T-construction for the cubical simulations.

For each of the two spin domains, we compute the Euler characteristic, Betti numbers, and the numbers of cells of each dimension.

To compute $\ec$ efficiently in the cubical case, we use a local lookup-table method based on overlapping $2\times2$ neighborhoods of spins. There are $2^4=16$ possible spin configurations in such a neighborhood, and a local Euler contribution is precomputed for each of them. Because cells of different dimensions occur with different multiplicities when the overlapping neighborhoods are summed over the periodic lattice, the corresponding dimension-dependent weights are incorporated directly into the lookup-table entries. With this convention, the global Euler characteristic is obtained as
\begin{equation}
\ec
=
\frac{1}{4}
\sum_{Q\in\mathcal{Q}}
\ec_{\mathrm{LUT}}(Q),
\label{eq:ec_lookup}
\end{equation}
where $\mathcal{Q}$ denotes the set of all overlapping $2\times2$ neighborhoods, with periodic boundary conditions, and $\ec_{\mathrm{LUT}}(Q)$ is the weighted local contribution associated with the spin configuration in $Q$.

\subsection{Reweighting}
For each observable $O$, the multihistogram reweighting procedure~\cite{FerrenbergSwendsen1989} reconstructs the temperature-dependent expectation value $\langle O\rangle_T$ and the corresponding finite-volume susceptibility curve $\chi_O^{(L)}(T)$ over the scanned temperature interval. The reweighting is performed directly on the recorded energies of individual configurations, without energy binning. The specific fluctuation estimator used for the $Z_2$-odd observables considered in the finite-size analysis is given in Sec.~\ref{sec:fss}.

For each lattice size $L$, we define the pseudocritical temperature as the location of the maximum of the susceptibility,
\begin{equation}
T_O^c(L)=\arg\max_T \chi_O^{(L)}(T).
\label{eq:pseudocritical_temperature}
\end{equation}
Statistical uncertainties are estimated using a circular moving block bootstrap, as described in Sec.~\ref{sec:fss}. The complete multihistogram reconstruction is repeated for $B=200$ bootstrap replicas, and the uncertainty of $T_O^c(L)$ is obtained from the standard deviation of the corresponding bootstrap distribution.

Once the pseudocritical temperatures are determined for all system sizes, we perform a finite-size scaling analysis. In particular, we fit their shift according to
\begin{equation}
T_O^c(L)=T_c(\infty)+A\,L^{-1/\nu},
\label{eq:fit_Tc_shift}
\end{equation}
to extract $T_c(\infty)$ and the critical exponent $\nu$.

OpenAI's ChatGPT (GPT-5.6 and GPT-6) was used to assist with reviewing and debugging the Monte Carlo simulation and multihistogram codes, including checks related to autocorrelations
and the implementation of the circular moving block bootstrap. It was also used to assist with developing and refining Python scripts for processing simulation outputs, numerical analysis, and figure generation. The authors specified the computational procedures, reviewed the AI-assisted code modifications, and examined the resulting numerical outputs and diagnostics.

\section{Results}\label{sec:results}

In this section, we analyze numerical results obtained from Monte Carlo simulations of the Ising model on regular square (2D) and triangular (2D) lattices with periodic boundary conditions. 

Ensembles of configurations were obtained for a set of values of $L$ and $T$ using Monte Carlo methods. A combination of the Metropolis-Hastings (MH) and Wolff update algorithms was used. Starting from a random spin configuration, the system was first thermalized by performing 2000 hybrid update steps. Each hybrid update step consisted of one complete MH sweep over all spins of the lattice, followed by 15 Wolff cluster updates. After thermalization, 200 configurations were recorded for each lattice size and temperature, with successive recorded configurations separated by $N_{\mathrm{decorr}}=10$ hybrid update steps.
The simulations were performed for several lattice sizes. Unless otherwise stated, the presented data correspond to $L = 3072$ for both lattice types.

For the square lattice, the temperature range spans from $T = 2.22$ to $T =
2.32$ with a step of $\Delta T = 0.0005$. 
For the triangular lattice, the temperature range extends from $T = 3.62$ to $T
= 3.72$ with the same step size. In all cases, simulations were initialized with the same random seed (123456) to ensure reproducibility. We additionally examined cross-temperature and
cross-size correlations between the recorded Monte Carlo time series and found no systematic evidence of statistically significant linear dependence.

Each configuration consists of spin variables $s_i \in \{-1,1\}$. For each configuration, the corresponding cubical (or simplicial, in the triangular case) complexes were constructed separately for the positive and negative spin domains.
\subsection{Validation}
\begin{figure*}[t]
\centering
\begin{minipage}{0.9\textwidth}
  \centering
  \includegraphics[width=0.48\linewidth]{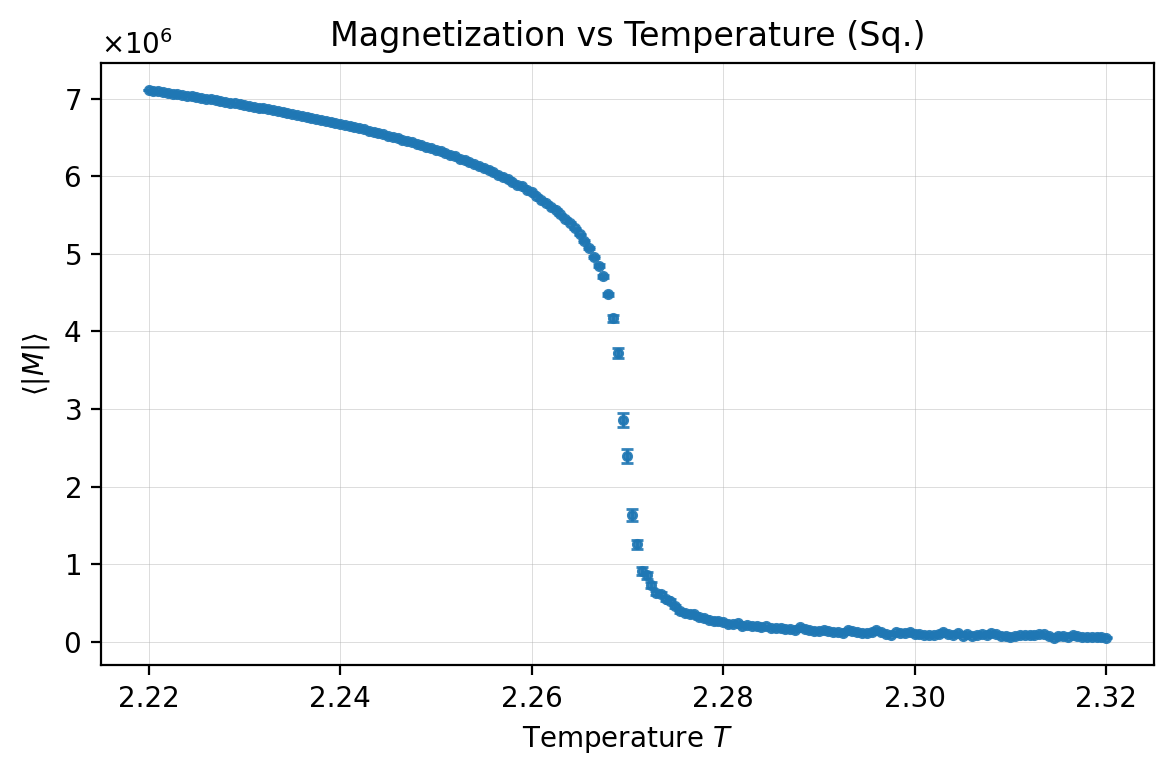}
  \includegraphics[width=0.48\linewidth]{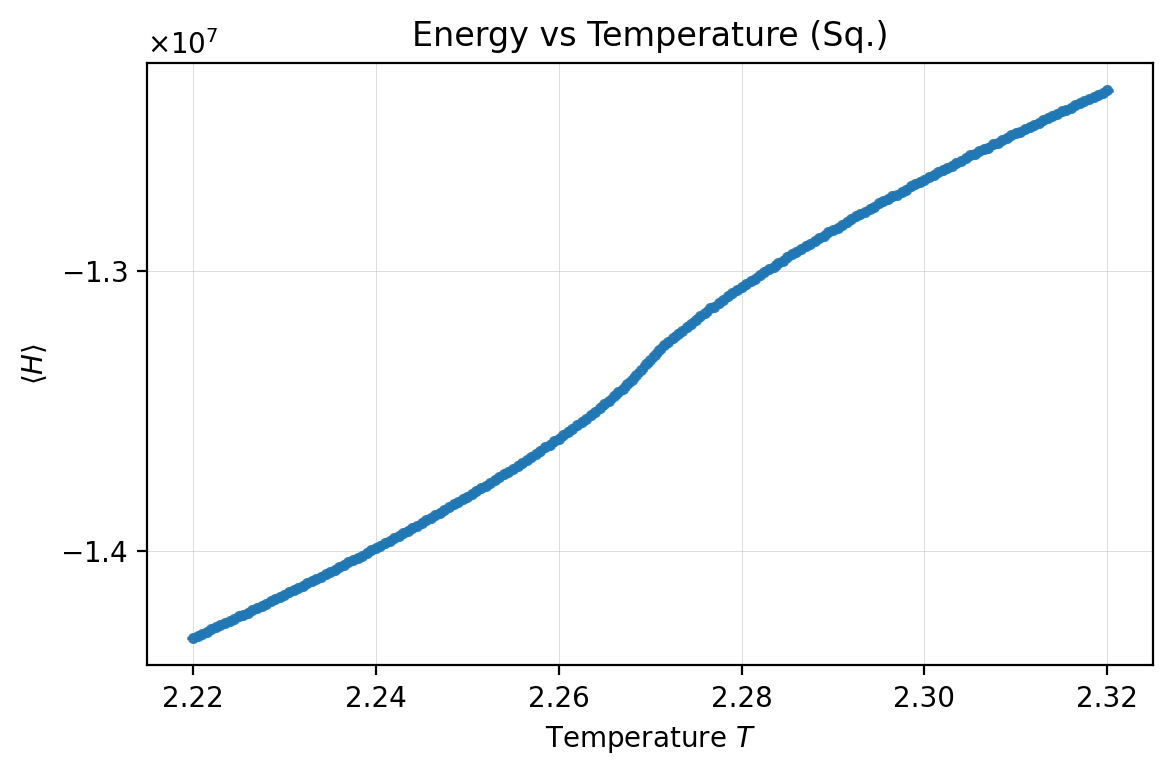}\\[0.7em]

  \caption{Magnetization $\langle |M| \rangle$ (left column) and mean energy
    $\langle H \rangle$ (right column) as functions of temperature for the two-dimensional square-lattice Ising model with periodic boundary conditions ($L=3072$, 200 samples per temperature). Points show sample means at each temperature, and error bars denote block-jackknife standard errors estimated from the recorded Monte Carlo time series. The drop of $\langle |M| \rangle$ and the change in slope of $\langle H \rangle$ near $T_c$ are consistent with the transition from the ordered to the disordered phase.
}
  \label{fig:thermo_ref}
\end{minipage}
\end{figure*}

To validate the numerical implementation and to provide a reference for the
forthcoming topological analysis, we report the raw thermodynamic observables of
the Ising model near the critical region. Figure~\ref{fig:thermo_ref} shows the
magnetization and total energy values obtained from the recorded configurations at each temperature for the square lattice with periodic
boundary conditions. The sharp loss of magnetization near the critical temperature clearly signals the transition from the ferromagnetic to the paramagnetic phase, while the change in the total energy curve signals the approach to the critical temperature.
These reference plots illustrate the expected physical behavior and will serve as benchmarks for the topological quantities discussed later.

\subsection{Euler characteristics}
\begin{figure*}[t]
\centering
\includegraphics[width=0.47\textwidth]{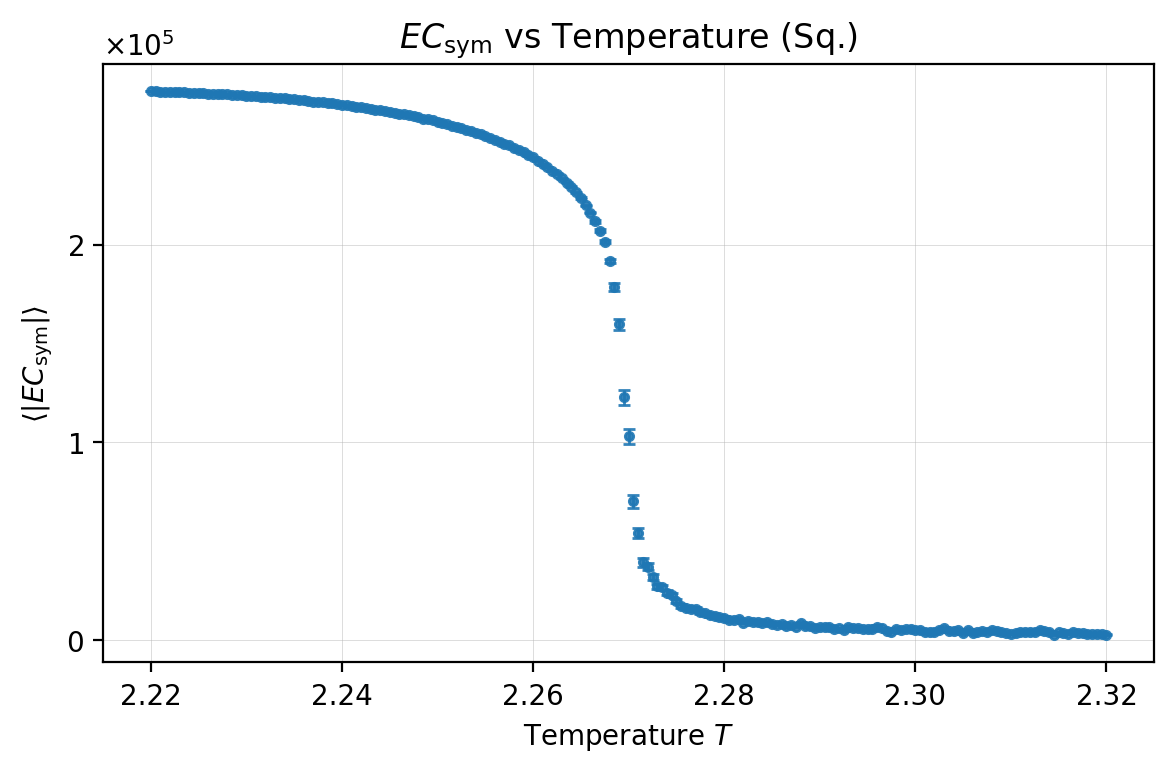}
\includegraphics[width=0.47\textwidth]{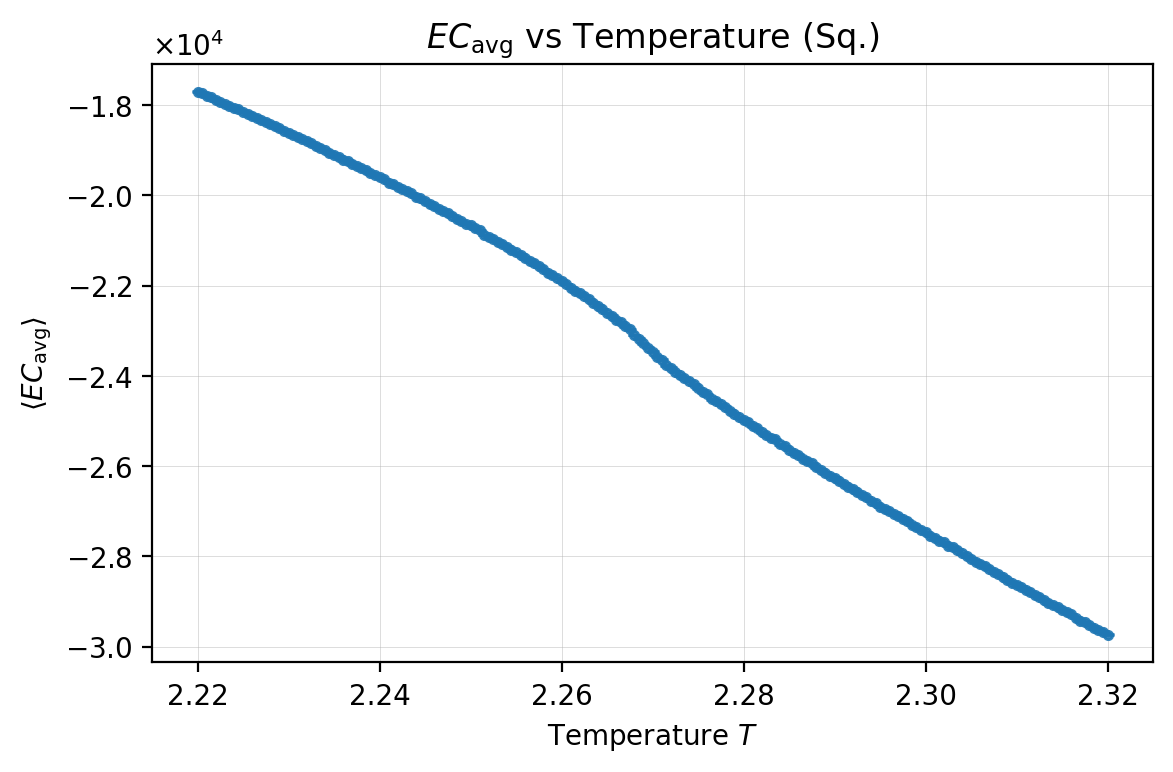}\\[0.7em]
\caption{
    Symmetrized $\langle | \ecs | \rangle$ (left column) and averaged $\langle \eca \rangle$ (right column) Euler characteristics as functions of temperature for the two-dimensional square-lattice Ising model with periodic boundary conditions ($L=3072$, 200 samples per temperature). Points show sample means at each temperature, and error bars denote block-jackknife standard errors estimated from the recorded Monte Carlo time series. The behavior of $\langle | \ecs | \rangle$ parallels that of the magnetization, while $\langle \eca \rangle$ mirrors the total energy (up to a sign), suggesting a close connection between the Euler observables and the conventional thermodynamic observables $M$ and $H$, which is clarified analytically below.
}
\label{fig:euler_vs_T}
\end{figure*}

We now examine the temperature dependence of the symmetrized and averaged Euler characteristics shown in Figure~\ref{fig:euler_vs_T}. The plotted quantities are not normalized by the number of spins, whereas the corresponding per-spin quantities, $\ecsp$ and $\ecap$, are used in the finite-size analysis below. The mean absolute symmetrized Euler characteristic $\langle|\ecs|\rangle$ follows a trend similar to the mean absolute magnetization $\langle|M|\rangle$: it decreases smoothly with temperature, with a rapid drop near the critical region. This resemblance suggests that $\ecs$ encodes information related to the macroscopic imbalance between the two spin phases through the topology of the spin domains. However, this analogy is restricted to the vicinity of the transition and does not extend to the zero-temperature limit, where $\ecs$ vanishes despite the magnetization reaching its maximal magnitude, as discussed below.

The averaged Euler characteristic $\eca$ exhibits a different behavior. It decreases monotonically with temperature, showing an inflection-like change near $T_c$. Its overall shape resembles the total energy, up to a change of sign. This resemblance is later explained by an exact relation between $\eca$ and the energy.

In the following analysis, we focus mainly on $\ecs$. The reason is that $\ecs$ behaves as an order-parameter-like observable, it is sensitive to the imbalance between positive and negative spin domains and becomes small in the disordered phase. This makes it a natural candidate for testing whether a topological observable of the spin-domain structure can reproduce the critical behavior usually detected through the magnetization.

We therefore use $\ecs$ as the main topological observable in the finite-size analysis below. In particular, we investigate whether the pseudocritical temperatures extracted from $\ecs$ are compatible with the known critical behavior of the Ising model.

\subsection{Relation between the symmetrized Euler characteristic and magnetization}

The similarity between the symmetrized Euler characteristic and the magnetization observed above can be understood more directly through exact relations between the two quantities. We derive such relations separately for the square and triangular lattices. In both cases, the symmetrized Euler characteristic can be expressed as a contribution proportional to the magnetization together with a local geometric correction measuring the imbalance between fully positive and fully negative local spin configurations.

\subsubsection{Square lattice}
The similarity between the symmetrized Euler characteristic and the magnetization observed above can be understood more directly through an exact relation between the two quantities. We derive this relation using the T-construction. Since $\ecs$ is independent of the choice between the V- and T-constructions, the resulting value of $\ecs$ is the same in both cubical representations.

Consider the two-dimensional square lattice. In the T-construction, let $F_k^\pm$ denote the number of $k$-dimensional cells in the positive and negative complexes, respectively. Since every spin is represented by a two-dimensional cell,
\begin{equation}
F_2^+-F_2^-=M,
\label{eq:magnetization_from_faces}
\end{equation}
where $M=\sum_i s_i$ denotes the total magnetization.

To express the lower-dimensional cell counts, we classify vertices and edges according to their local spin environment. A vertex of the square lattice is incident on four two-dimensional cells. We denote by $F_0^{+4}$ and $F_0^{-4}$ the numbers of vertices for which all four incident cells carry positive or negative spins, respectively. We denote by $F_0^{+2}$ and $F_0^{-2}$ the numbers of vertices with local spin imbalance $+2$ and $-2$, corresponding to configurations with three spins of one sign and one spin of the opposite sign. Finally, $F_0^0$ denotes vertices with two positive and two negative incident cells.

Accordingly, the total numbers of vertices belonging to the positive and negative complexes are
\begin{align}
F_0^+ &= L^2-F_0^{-4},
\label{eq:positive_vertex_decomposition}\\
F_0^- &= L^2-F_0^{+4}.
\label{eq:negative_vertex_decomposition}
\end{align}
Hence,
\begin{equation}
F_0^+-F_0^-=
F_0^{+4}-F_0^{-4}
\label{eq:vertex_difference_local_classes}
\end{equation}

We classify edges in an analogous way. Each edge is shared by two two-dimensional cells. Let $F_1^{+2}$ denote the number of edges for which both incident cells carry positive spins, and let $F_1^{-2}$ denote the corresponding number for negative spins. An edge shared by one positive and one negative cell belongs to both complexes and therefore cancels in the difference $$F_1^+-F_1^-.$$ Hence only edges whose two incident cells have the same sign contribute to this difference. Therefore,
\begin{equation}
F_1^+-F_1^-=
F_1^{+2}-F_1^{-2}.
\label{eq:edge_difference_local_classes}
\end{equation}

Using
\begin{equation}
2\ecs=\ec^+-\ec^-,
\label{eq:ecsym_difference}
\end{equation}
together with
\begin{equation}
\ec^\pm=F_0^\pm-F_1^\pm+F_2^\pm,
\end{equation}
we obtain
\begin{align}
2\ecs
&=
\left(F_0^+-F_0^-\right)
-
\left(F_1^+-F_1^-\right)
+
\left(F_2^+-F_2^-\right)
\nonumber\\
&=
M
+
\left(F_0^{+4}-F_0^{-4}\right)
-
\left(F_1^{+2}-F_1^{-2}\right).
\label{eq:ecsym_local_decomposition}
\end{align}

To eliminate the edge contribution in Eq.~\eqref{eq:ecsym_local_decomposition}, we express the difference $F_1^{+2}-F_1^{-2}$ in terms of the local vertex classes. Since each edge is shared by two neighboring $2\times2$ vertex environments, the corresponding edge count appears twice in the incidence sum. A vertex of type $+4$ contributes four $+2$ edges, while a vertex of type $+2$ contributes two; the negative configurations contribute analogously. The balanced $F_0^0$ configurations give equal contributions to the positive and negative edge counts and therefore cancel in the difference. Hence,
\begin{equation}
2\left(F_1^{+2}-F_1^{-2}\right)
=
4\left(F_0^{+4}-F_0^{-4}\right)
+
2\left(F_0^{+2}-F_0^{-2}\right).
\label{eq:edge_difference_from_vertices}
\end{equation}

Multiplying Eq.~\eqref{eq:ecsym_local_decomposition} by two and substituting Eq.~\eqref{eq:edge_difference_from_vertices}, we obtain
\begin{equation}
\begin{split}
4\ecs
&=
2M
+
2\left(F_0^{+4}-F_0^{-4}\right)
-
2\left(F_1^{+2}-F_1^{-2}\right)
\\
&=
2M
-
2\left(F_0^{+4}-F_0^{-4}\right)
-
2\left(F_0^{+2}-F_0^{-2}\right).
\end{split}
\label{eq:ecsym_after_edge_elimination}
\end{equation}

The remaining vertex contribution can be related directly to the magnetization. Summing the local spin imbalance over all vertex environments counts each spin four times, since every two-dimensional cell is incident on four vertices. A vertex of type $+4$ contributes $+4$ to this sum, a vertex of type $+2$ contributes $+2$, and a balanced vertex contributes zero; the corresponding negative configurations contribute $-4$ and $-2$, respectively. Therefore,
\begin{equation}
4M
=
4\left(F_0^{+4}-F_0^{-4}\right)
+
2\left(F_0^{+2}-F_0^{-2}\right).
\label{eq:magnetization_from_vertex_classes}
\end{equation}
Using Eq.~\eqref{eq:magnetization_from_vertex_classes} to eliminate
$F_0^{+2}-F_0^{-2}$ from Eq.~\eqref{eq:ecsym_after_edge_elimination}, we obtain
\begin{equation}
4\ecs
=
-2M
+
2\left(F_0^{+4}-F_0^{-4}\right),
\label{eq:ecsym_magnetization_intermediate}
\end{equation}
and hence
\begin{equation}
\ecs 
=-\frac{M}{2}+\frac{
F_0^{+4}-F_0^{-4}
} {2}.\label{eq:ecsym_magnetization_relation}
\end{equation}

or
\begin{equation}
M
=
-2\ecs
+
\left(F_0^{+4}-F_0^{-4}\right).\label{eq:magnetization_decomposition}
\end{equation}

Equation~\eqref{eq:magnetization_decomposition} gives an exact decomposition of the magnetization into a contribution proportional to the symmetrized Euler characteristic and a local geometric contribution. The latter measures the imbalance between vertices lying entirely inside positive and negative spin domains, since a vertex contributes to $F_0^{+4}$ or $F_0^{-4}$ only when all four incident cells belong to the same spin domain.

Also, combining Eqs.~\eqref{eq:magnetization_from_vertex_classes} and~\eqref{eq:ecsym_magnetization_relation} yields
\begin{equation}
    \ecs=\frac{F_0^{-2}-F_0^{+2}}{4}.
\end{equation}
Thus, only vertex environments containing three spins of one sign and one of the opposite sign contribute to this representation: those with three negative spins contribute $+1/4$, while those with three positive spins contribute $-1/4$.
\subsubsection{Triangular lattice}

For the triangular lattice, we consider the simplicial construction introduced above. Let $F_k^\pm$ denote the numbers of $k$-simplices in the positive and negative complexes, respectively. In this construction, lattice sites are the $0$-simplices, nearest-neighbor pairs of equal sign form $1$-simplices, and an elementary triangle forms a $2$-simplex when all three of its vertices carry the same spin.

Using
\begin{equation}
    \begin{aligned}
        2\ecs
        &=
        \ec^+ - \ec^-
        \\
        &=
        \left(F_0^+ - F_0^-\right)
        -
        \left(F_1^+ - F_1^-\right)
        +
        \left(F_2^+ - F_2^-\right).
    \end{aligned}
    \label{eq:ecsym_triangular_decomposition}
\end{equation}
we consider the three contributions separately.

Since every lattice site corresponds directly to one spin,
\begin{equation}
F_0^+ - F_0^- = M.
\label{eq:triangular_vertex_difference}
\end{equation}

The difference between the numbers of positive and negative edges can also be expressed directly in terms of the magnetization. For a nearest-neighbor pair $\langle ij\rangle$, the contribution to $F_1^+-F_1^-$ is $+1$ if both spins are positive, $-1$ if both spins are negative, and $0$ if the pair contains spins of opposite sign. This contribution can be written compactly as
\begin{equation}
\frac{s_i+s_j}{2}.
\end{equation}
Summing over all nearest-neighbor edges gives
\begin{equation}
F_1^+ - F_1^-
=
\frac{1}{2}
\sum_{\langle ij\rangle}
\left(s_i+s_j\right).
\end{equation}
Each site of the triangular lattice has six nearest neighbors, so each spin appears six times in the incidence sum. Therefore,
\begin{equation}
F_1^+ - F_1^- = 3M.
\label{eq:triangular_edge_difference}
\end{equation}

Substituting Eqs.~\eqref{eq:triangular_vertex_difference} and
\eqref{eq:triangular_edge_difference} into
Eq.~\eqref{eq:ecsym_triangular_decomposition}, we obtain
\begin{equation}
2\ecs
=
-2M
+
\left(F_2^+ - F_2^-\right).
\label{eq:triangular_ecsym_relation}
\end{equation}
Equivalently,
\begin{equation}
\ecs
=
-M
+
\frac{1}{2}
\left(F_2^+ - F_2^-\right),
\label{eq:triangular_ecsym_magnetization_relation}
\end{equation}
or
\begin{equation}
M
=
-\ecs
+
\frac{1}{2}
\left(F_2^+ - F_2^-\right).
\label{eq:triangular_magnetization_decomposition}
\end{equation}

Equation~\eqref{eq:triangular_magnetization_decomposition} has the same qualitative structure as the corresponding relation for the square lattice. The additional term measures the imbalance between fully positive and fully negative local simplices. Here, $F_2^+$ and $F_2^-$ count elementary triangles whose three vertices carry positive or negative spins, respectively.\newline

Taken together, these relations suggest a common interpretation for both lattice geometries. In both decompositions, $\ecs$ contains a term proportional to the magnetization with a negative coefficient, together with a local geometric correction; their relative signs therefore depend on the balance between these two contributions. Equivalently, the magnetization can be decomposed into a contribution determined by $\ecs$ and an additional local geometric contribution. The latter measures the imbalance between structures lying in the interior of positive and negative spin domains: vertices surrounded by four equal spins on the square lattice, and fully occupied elementary triangles on the triangular lattice. These terms therefore probe the local bulk, or thickness, of the corresponding spin domains. In this sense, the magnetization contains both a topological contribution, encoded by $\ecs$, and a geometric contribution associated with the interior structure of the domains.

The order-parameter-like behavior of $\ecs$ near the critical point does not imply the same temperature dependence as the magnetization throughout the ordered phase. In particular, in the zero-temperature limit, all spins are aligned and one of the spin complexes covers the entire torus, whose Euler characteristic vanishes, while the other complex is empty. Consequently, $\ecs=0$ for both lattice geometries, even though the absolute magnetization reaches its maximum value $|M|=N_s$.

Beyond this geometric interpretation, the exact representation shows that $\ecs$ is a local $Z_2$-odd lattice observable, placing it in the same symmetry sector as the magnetization. The implications of this observation for its critical scaling are discussed below.

\subsection{Relation between the averaged Euler characteristic and energy}

A similar relation can be obtained for the averaged Euler characteristic. Here we restrict attention to the square lattice, since this observable vanishes identically on the triangular lattice. Using the T-construction, Eq.~\eqref{eq:eca_T}, and the identifications $N_+=F_0^{+4}$ and $N_-=F_0^{-4}$, we have
\begin{equation}
2\eca^T=L^2-\left(F_0^{+4}+F_0^{-4}\right)-E_{+-}.
\label{eq:ecavg_local_decomposition}
\end{equation}

The energy can be expressed directly in terms of $E_{+-}$. On the two-dimensional square lattice with periodic boundary conditions there are $2L^2$ nearest-neighbour edges. With the convention used here,
\begin{equation}
H=-2L^2+2E_{+-},
\label{eq:energy_from_mixed_edges}
\end{equation}
and therefore
\begin{equation}
E_{+-}=L^2+\frac{H}{2}.
\label{eq:mixed_edges_from_energy}
\end{equation}

Substituting Eq.~\eqref{eq:mixed_edges_from_energy} into Eq.~\eqref{eq:ecavg_local_decomposition}, we obtain
\begin{align}
2\eca^T
&=L^2-\left(F_0^{+4}+F_0^{-4}\right)-\left(L^2+\frac{H}{2}\right)
\nonumber\\
&=-\frac{H}{2}-\left(F_0^{+4}+F_0^{-4}\right).
\end{align}
Hence,
\begin{equation}
\eca^T=-\frac{H}{4}-\frac{F_0^{+4}+F_0^{-4}}{2}.
\label{eq:eca_as_H}
\end{equation}
Equivalently,
\begin{equation}
H=-4\eca^T-2\left(F_0^{+4}+F_0^{-4}\right).
\label{eq:H_decomp}
\end{equation}

Equation~\eqref{eq:H_decomp} expresses the energy as a contribution proportional to the averaged Euler characteristic together with a local geometric term counting vertices whose four incident two-dimensional cells belong entirely to a single spin domain, irrespective of its sign. 

This relation also gives a simple interpretation of $\eca^T$ in terms of local spin configurations. Around each vertex, configurations with three spins of one sign and one of the opposite sign, as well as configurations with two positive and two negative spins with equal signs adjacent in cyclic order, contain two equal-spin and two opposite-spin pairs, whose energy contributions cancel. The four surrounding pairs contribute $-4$ when all spins are equal and $+4$ when they form a checkerboard pattern. Since each nearest-neighbor bond is counted around two vertices, the total energy is
\begin{equation}
    H=-2\left(F_0^{+4}+F_0^{-4}\right)+2F_0^C,
\end{equation}
where $F_0^C$ counts vertices surrounded by a checkerboard pattern and satisfies $F_0^C\leq F_0^0$. Substituting this expression into Eq.~\eqref{eq:eca_as_H}, the contributions from vertices surrounded by four equal spins cancel, leaving
\begin{equation}
    \eca^T=-\frac{F_0^C}{2}.
    \label{ec_avg:checkerboard}
\end{equation}

\subsection{Implications for the averaged Euler characteristic in the low-temperature phase}

The relation derived above allows us to revisit a conjecture proposed by Blanchard et al.~\cite{BlanchardEtAl2006} concerning the mean Euler characteristic of the two-dimensional Ising model. For the standard Ising case ($Q=1$), their global Euler characteristic corresponds to the sum of the Euler characteristics of the positive and negative spin complexes, and thus to twice our averaged Euler characteristic in the V-construction. Using Eq.~\eqref{ec_avg:odd} together with Eq.~\eqref{ec_avg:checkerboard}, we obtain
\begin{equation}
    \chi(\sigma)=2\eca^V=F_0^C,
    \label{eq:blanchard_checkerboard}
\end{equation}
where $F_0^C$ counts elementary plaquettes whose four spins form a checkerboard pattern. For periodic boundary conditions, this identity holds for every spin configuration.

The mean Euler characteristic per site considered in Ref.~\cite{BlanchardEtAl2006}, denoted there by $\chi_\Lambda(\beta)$, can be expressed in our notation as
\begin{equation}
\ec_L(T)=\frac{2\langle\eca^V\rangle_{L,T}}{L^2}=\frac{\langle F_0^C\rangle_{L,T}}{L^2}=P_{L,T}(C_p),
\label{eq:blanchard_checkerboard_probability}
\end{equation}
where $C_p$ denotes the event that the four spins at the vertices of a fixed elementary plaquette $p$ form a checkerboard pattern.

We can establish a strictly positive lower bound on this probability. For the ferromagnetic Ising model with $J=1$ and zero external field, each spin has four nearest neighbors. The conditional probability of either prescribed spin value, given all other spins, is therefore bounded below by
\begin{equation}
p_{\min}(T)=\frac{1}{1+e^{8/T}}.
\end{equation}
By applying the chain rule for conditional probabilities, the probability of any prescribed configuration of four spins is at least $p_{\min}^4$. Since there are two distinct checkerboard configurations, we obtain
\begin{equation}
\ec_L(T)\geq\frac{2}{(1+e^{8/T})^4}>0.
\label{eq:blanchard_positive_bound}
\end{equation}

Importantly, this lower bound is independent of the lattice size. Consequently,
\begin{equation}
\liminf_{L\to\infty}\ec_L(T)\geq\frac{2}{(1+e^{8/T})^4}>0
\label{eq:blanchard_thermodynamic_bound}
\end{equation}
for every positive temperature $T$. For finite square lattices with open boundaries, the relation between the Euler characteristic and the checkerboard count contains additional boundary contributions of order $O(L)$. These vanish after normalization by $L^2$ in the thermodynamic limit, leaving the conclusion unchanged.

Blanchard et al.~\cite{BlanchardEtAl2006} conjectured that the mean Euler characteristic per site vanishes in the thermodynamic limit for $T<T_c$. Our result shows that exact vanishing is not possible for the standard Ising model at any positive temperature. Although the checkerboard density becomes very small at low temperatures, it remains strictly positive. This does not exclude its vanishing in the zero-temperature limit $T\to 0$.

\subsection{Pseudocritical temperatures and finite-size scaling}\label{sec:fss}
Having established that the topological observables display temperature dependences similar to standard thermodynamic quantities, we now turn to a quantitative finite-size analysis. In this section we use the multihistogram method to interpolate both physical and topological observables between the simulated temperatures, identify their pseudocritical points, and analyze their scaling with the system size. The magnetization is used as a benchmark for the procedure, while $\ecs$ is used to test whether a topological observable extracted from the spin-domain geometry can locate the phase transition quantitatively.

Motivated by the exact relations derived above, we additionally consider the local domain-imbalance contribution $D$, defined for the two lattice geometries as
\begin{equation}
D=
\begin{cases}
F_0^{+4}-F_0^{-4}, & \text{square lattice},\\[2mm]
\dfrac{1}{2}\left(F_2^+-F_2^-\right), & \text{triangular lattice}.
\end{cases}
\label{eq:D_definition}
\end{equation}
With this definition, the exact decompositions take the form
\begin{equation}
M=-2\ecs+D
\end{equation}
for the square lattice and
\begin{equation}
M=-\ecs+D
\end{equation}
for the triangular lattice. We also introduce the corresponding quantity per lattice site,
\begin{equation}
d=\frac{D}{N_s}.
\label{eq:d_definition}
\end{equation}
Its finite-size behavior allows us to determine how the critical information contained in the magnetization is distributed between the topological contribution $\ecsp$ and the remaining local geometric term $d$.

We first validate the procedure using the magnetization and then apply the same analysis to $\ecsp$ and $d$.

For the $Z_2$-odd observables considered in the finite-size analysis, $O=m,\ecsp,d$, we use the finite-volume fluctuation estimator
\begin{equation}
\chi_O^{(L)}
=
\beta N_s
\left(
\langle O^2\rangle
-
\langle |O|\rangle^2
\right).
\label{eq:numerical_susceptibility}
\end{equation}
In the multihistogram calculation, the absolute value of each
observable is stored configuration by configuration. Since $|O|^2=O^2$, multiplying the variance of the stored quantity by $N_s$ gives
\begin{equation}
    N_s
    \left(
    \langle O^2\rangle
    -
    \langle |O|\rangle^2
    \right),
\end{equation}
while the factor $\beta=1/T$ is included in the subsequent analysis.

For each lattice size $L$ and each observable $O$, we use multihistogram reweighting~\cite{FerrenbergSwendsen1989}to reconstruct the finite-volume susceptibility curve $\chi_O^{(L)}(T)$ over the scanned temperature interval. The pseudocritical temperature is identified with the position of its maximum.

To assess the reliability of the multihistogram reconstruction, we examined the overlap between energy distributions at adjacent simulation temperatures. For this diagnostic, the energy distributions were represented by normalized histograms with 25 equally spaced bins over a common energy range for each pair of temperatures. On the square lattice, the minimum overlap decreases from 0.725 at $L=64$ to 0.415 at $L=3072$, while the median overlap decreases from 0.825 to 0.615. Thus, although the overlap decreases with increasing system size, the adjacent energy distributions retain appreciable overlap even for the largest systems.

Statistical uncertainties are estimated using a circular moving block bootstrap \cite{PolitisRomano1992} to account for residual autocorrelations between successive recorded configurations. For each lattice size and simulation temperature, we estimate the integrated autocorrelation time of seven time series:
\[
E,\quad |m|,\quad |\ecsp|,\quad |d|,
\quad m^2,\quad \ecsp^2,\quad d^2.
\]
For a time series $X$, the integrated autocorrelation time is estimated as
\begin{equation}
\tau_{\mathrm{int},X}(W)
=
\frac{1}{2}
+
\sum_{t=1}^{W}\rho_X(t),
\end{equation}
where $\rho_X(t)$ denotes the normalized autocorrelation function. The summation window $W$ is determined self-consistently as the first lag satisfying
$W>4\tau_{\mathrm{int},X}(W)$~\cite{Madras:1988ei}.
All autocorrelation times are expressed in units of recorded configurations.

The estimated integrated autocorrelation times were typically close to the independent-sample reference value of $\tau_{\mathrm{int}}=1/2$, indicating relatively weak residual temporal correlations between successive recorded configurations. For the square lattice, the largest estimated value was $1.10$, with no clear systematic dependence on lattice size. For the triangular lattice, the largest estimate reached $1.99$, with increased autocorrelation times of the energy observed for the largest lattice sizes.

For each lattice size $L$ and simulation temperature $T_i$, the bootstrap block length is determined from the largest estimated autocorrelation time across the seven time series,
\begin{equation}
\ell(L,T_i)
=
\left\lceil
10\max\left(\frac{1}{2},\tau_{\max}(L,T_i)\right)
\right\rceil.
\end{equation}

We generate $B=200$ bootstrap replicas. Within each replica, circular blocks are sampled with replacement independently for each simulated temperature until the original number of measurements at that temperature is recovered. The energy and all observables are resampled jointly using the same configuration indices, preserving their within-block correlations.

Although weak residual cross-temperature correlations cannot be entirely excluded, our analysis found no evidence of significant dependence. We therefore do not expect such correlations to materially affect the reported results.

For each bootstrap replica $b$, we perform a separate multihistogram reconstruction using the resampled measurements from all simulation temperatures at a given lattice size. The resulting susceptibility curve $\chi_{O,b}^{(L)}(T)$ is used to determine the pseudocritical temperature
\begin{equation}
T_{O,b}^c(L)
=
\arg\max_T \chi_{O,b}^{(L)}(T).
\label{eq:bootstrap_pseudocritical_temperature}
\end{equation}
The final pseudocritical-temperature estimate is the mean of the peak positions obtained from the bootstrap replicas,
\begin{equation}
\widehat{T}_O^c(L)
=
\frac{1}{B}
\sum_{b=1}^{B}T_{O,b}^c(L),
\end{equation}
and its statistical uncertainty is estimated by their sample standard deviation,
\begin{equation}
\sigma_{T_O^c(L)}
=
\sqrt{
\frac{1}{B-1}
\sum_{b=1}^{B}
\left(
T_{O,b}^c(L)-\widehat{T}_O^c(L)
\right)^2
}.
\end{equation}
The bootstrap procedure was performed with a fixed random seed of $123456$ to ensure reproducibility. The scripts used for bootstrap resampling, multihistogram reconstruction, and pseudocritical-temperature extraction are provided in the accompanying reproducibility repository.

\subsection{Fit window selection}

The finite-size scaling relation in Eq.~\eqref{eq:fit_Tc_shift} represents the leading asymptotic behavior as $L\to\infty$. For finite lattices, subleading corrections to scaling may be present, so it is important to verify whether the smallest simulated system sizes are already compatible with the leading-order scaling form. We therefore repeat the fits while progressively increasing the lower cutoff $L_{\min}$ and assess the goodness of fit. This procedure allows us to determine whether excluding smaller lattices is necessary.

For each observable $O$, we perform weighted fits of the pseudocritical temperatures to Eq.~\eqref{eq:fit_Tc_shift} for different choices of the lower cutoff $L_{\min}$. For each fit, we record the fitted parameters, the reduced $\chi^2$, and the goodness-of-fit probability
\begin{equation}
Q=P\left(\chi^2_{\mathrm{dof}}\geq\chi^2_{\mathrm{obs}}\right),
\end{equation}
where $\mathrm{dof}=n-k$ is the number of degrees of freedom, $n$ is the number of data points, and $k=3$ is the number of fitted parameters.

We select the smallest value of $L_{\min}$ for which $Q\geq 0.05$, corresponding to the fit that retains the largest number of system sizes while remaining statistically compatible with the leading-order scaling form. As additional consistency checks, we monitor the reduced $\chi^2$.

For all observables considered on the square lattice, this procedure selects the full fitting range beginning at $L_{\min}=64$. The fitting window is chosen without constraining the fits to reproduce the known Ising values of $T_c$ or $\nu$; agreement with theory is assessed only a posteriori.

\begin{table}[t]
\centering
\caption{Finite-size scaling results for the 2D square lattice, $L_{\min}=64$. The goodness-of-fit probability $Q$ is computed from the $\chi^2$ distribution with $n-3$ degrees of freedom.}
\label{tab:fss_Lmin64}

\scriptsize
\renewcommand{\arraystretch}{1.1}
\setlength{\tabcolsep}{3pt}

\begin{tabular}{l c
                S[table-format=1.6(2)]
                S[table-format=1.3(2)]
                S[table-format=1.3]
                S[table-format=1.3]}
\toprule
Observable & {$n$} & {$T_c(\infty)$} & {$\nu$} & {$\chi^2/\mathrm{dof}$} & {$Q$} \\
\midrule
$\ecsp$ & 12 & \num{2.269271(74)} & \num{0.934(15)} & \num{1.078} & \num{0.375} \\
$m$    & 12 & \num{2.269137(82)} & \num{1.005(18)} & \num{1.108} & \num{0.353} \\
$d$    & 12 & \num{2.269118(83)} & \num{1.014(19)} & \num{1.150} & \num{0.323} \\
\bottomrule
\end{tabular}
\end{table}

\begin{figure}[t]
    \centering
    \includegraphics[width=\columnwidth]{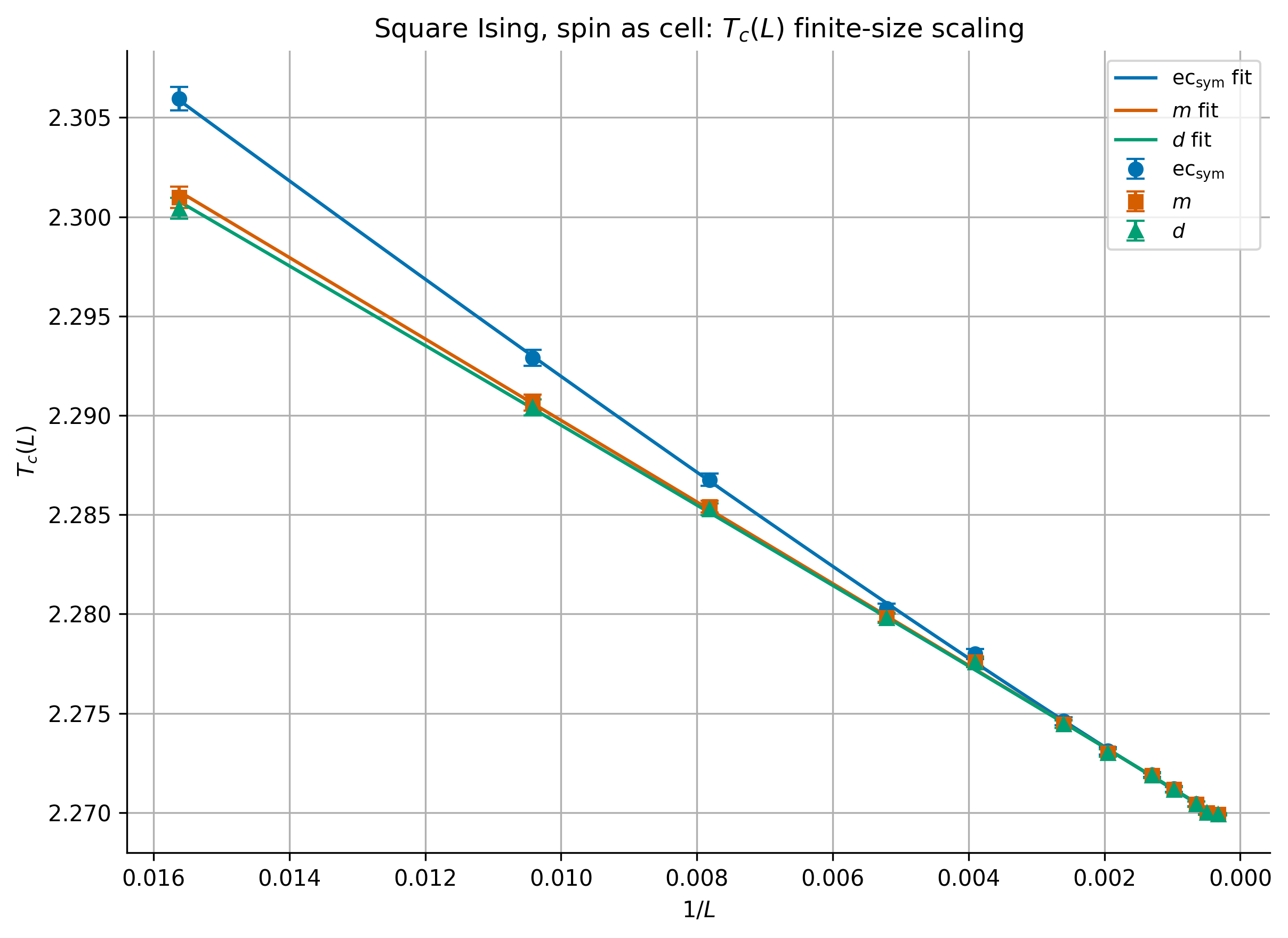}
    \caption{Finite-size scaling of the pseudocritical temperature $T_O^c(L)$ as a function of $1/L$ for magnetization, $\ecsp$ and $d$ on the two-dimensional square lattice, together with weighted fits to Eq.~\eqref{eq:fit_Tc_shift} in the selected fit window.}
    \label{fig:fss_square}
\end{figure}

As established above, $\ecsp$ and $d$ can both be expressed as local $Z_2$-odd lattice observables and are exactly related to the magnetization through the local decomposition derived above. They therefore belong to the same symmetry sector as $m$. Provided that they have a non-vanishing overlap with the leading odd scaling operator, standard renormalization-group arguments imply the same leading asymptotic scaling sector. In particular, the pseudocritical temperatures extracted from $m$, $\ecsp$, and $d$ are expected to approach $T_c$ with the same leading shift exponent $1/\nu$. Differences between the fitted values of $\nu$ obtained from these observables at finite lattice sizes are therefore naturally interpreted as systematics due to finite-size corrections rather than as evidence for different asymptotic critical behavior.

As visible in Fig.~\ref{fig:fss_square}, the pseudocritical temperatures obtained from all three observables approach the expected critical temperature as the system size increases. The magnetization provides the benchmark case. Its dependence on $1/L$ is nearly linear and the resulting fit gives an exponent close to the theoretical value $\nu=1$. The observable $d$ shows a very similar finite-size behavior and yields a value of $\nu$ consistent with that obtained from the magnetization. The topological observable $\ecsp$ follows the same overall convergence toward the critical point, but its dependence on $1/L$ is less linear over the full range of available sizes. This suggests stronger corrections to the leading finite-size scaling form and is reflected in the effective exponent $\nu_{\ecsp}=0.934(15)$ obtained from this fit, which deviates significantly from the theoretical value $\nu=1$, by approximately $4.4$ standard errors. The comparison is based on the fitting window selected using the goodness-of-fit criterion reported in Table~\ref{tab:fss_Lmin64}. The stronger curvature observed for $\ecsp$ makes it useful to examine more explicitly how the fitted parameters depend on the lower cutoff $L_{\min}$.
\subsection{Stability under changes of the fit window}
The comparison above is based on the fitting window selected using the goodness-of-fit criterion $Q\geq 0.05$. Since the topological observable shows a less linear dependence on $1/L$, we further examine how the fitted parameters depend on the range of lattice sizes included in the fit. We repeat the finite-size scaling fits while varying the lower cutoff $L_{\min}$ and keeping the upper cutoff $L_{\max}$ fixed. To check that the conclusions are not driven by the largest available lattice size alone, we perform this analysis for two choices of $L_{\max}$, namely the largest and the second-largest available sizes.

\begin{figure*}[t]
    \centering
    \includegraphics[width=0.48\linewidth]{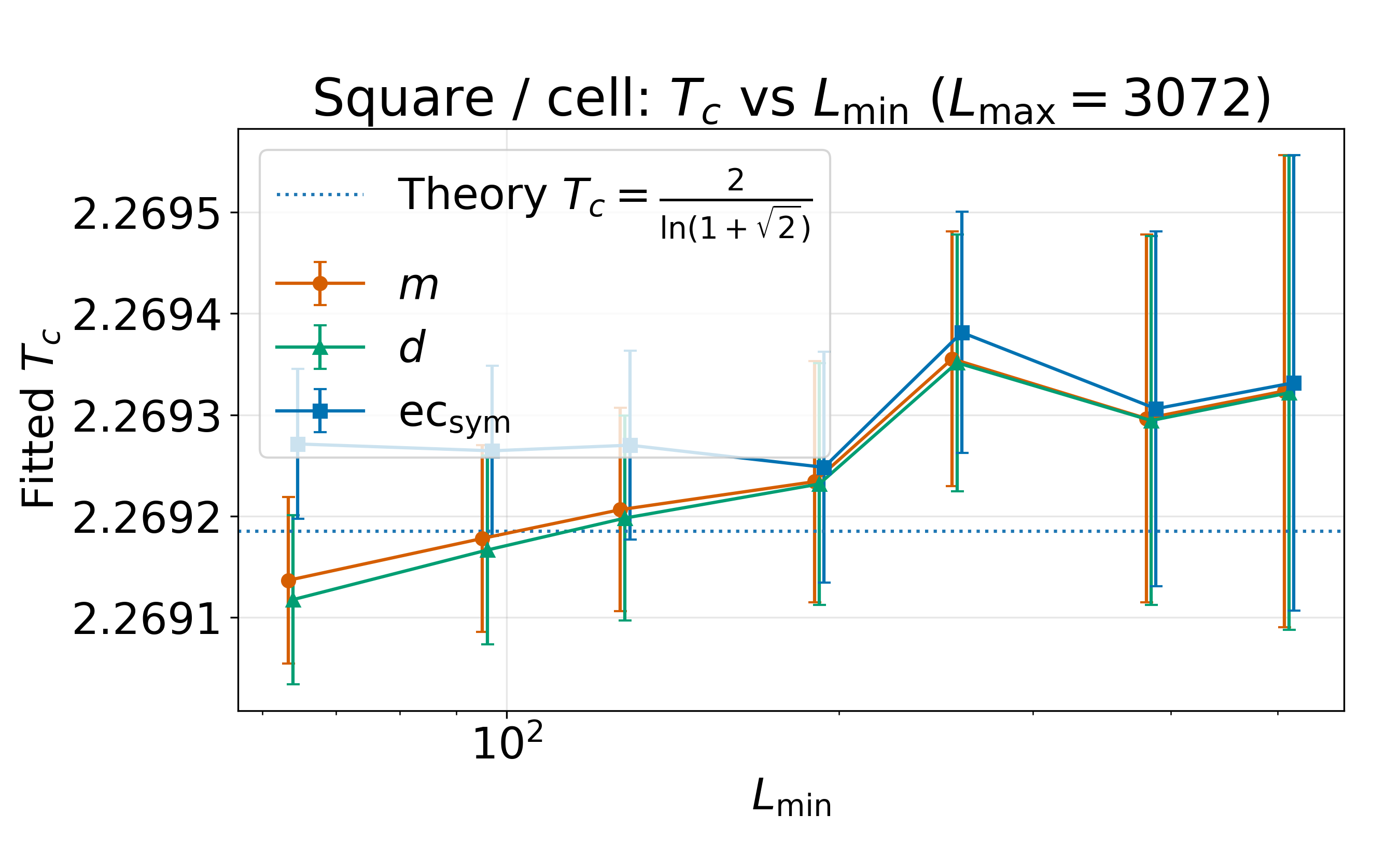}
    \includegraphics[width=0.48\linewidth]{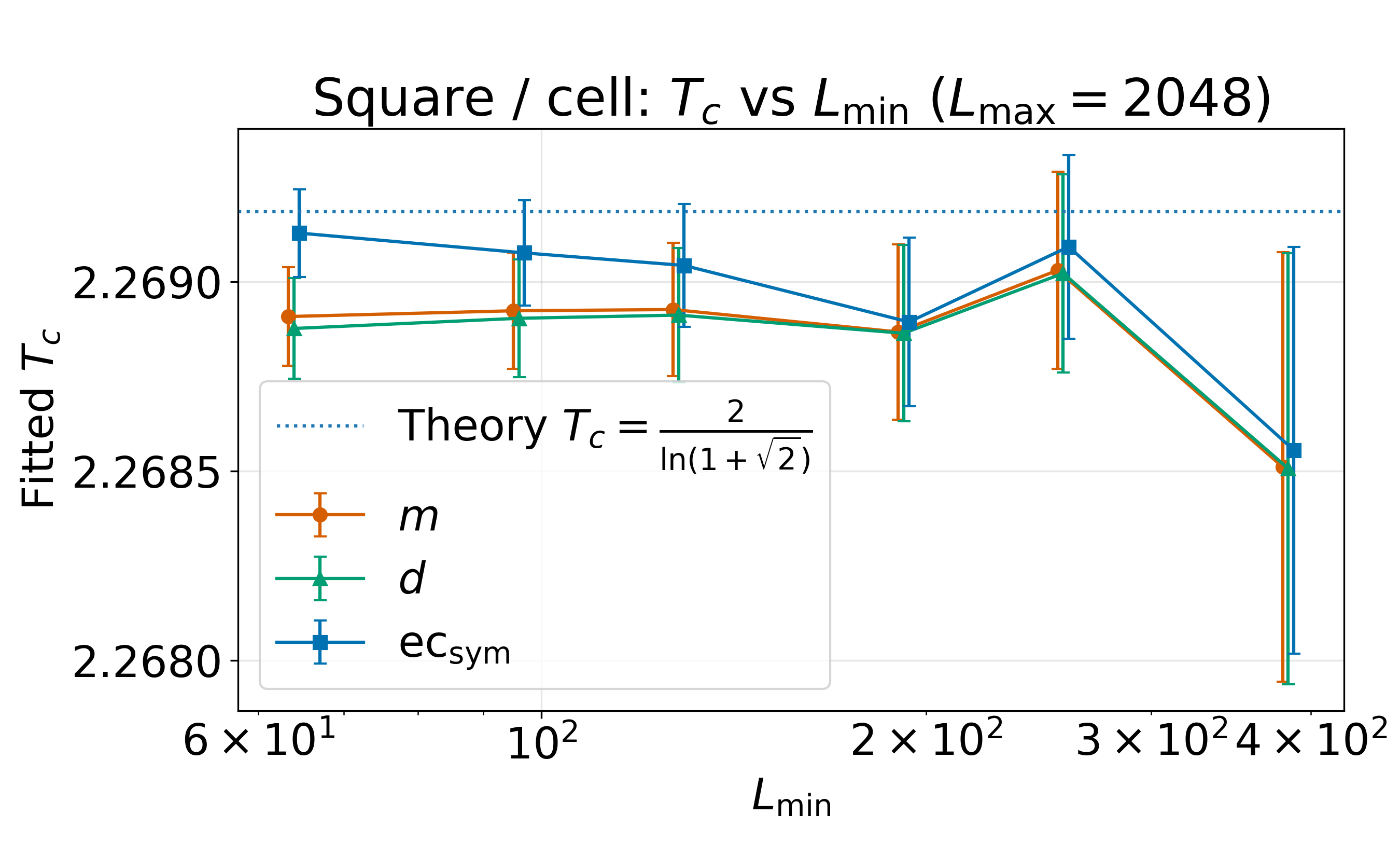}
    \includegraphics[width=0.48\linewidth]{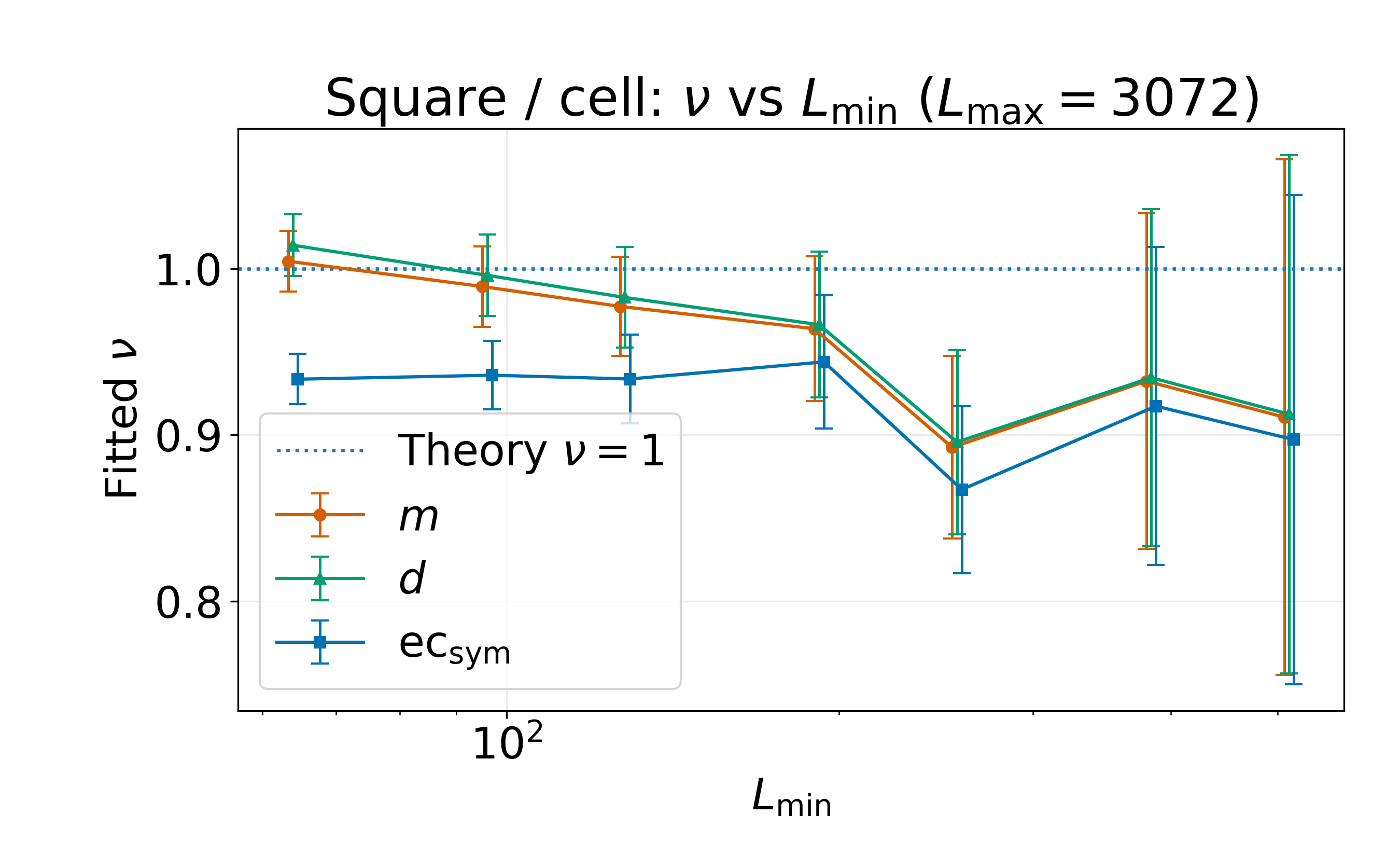}
    \includegraphics[width=0.48\linewidth]{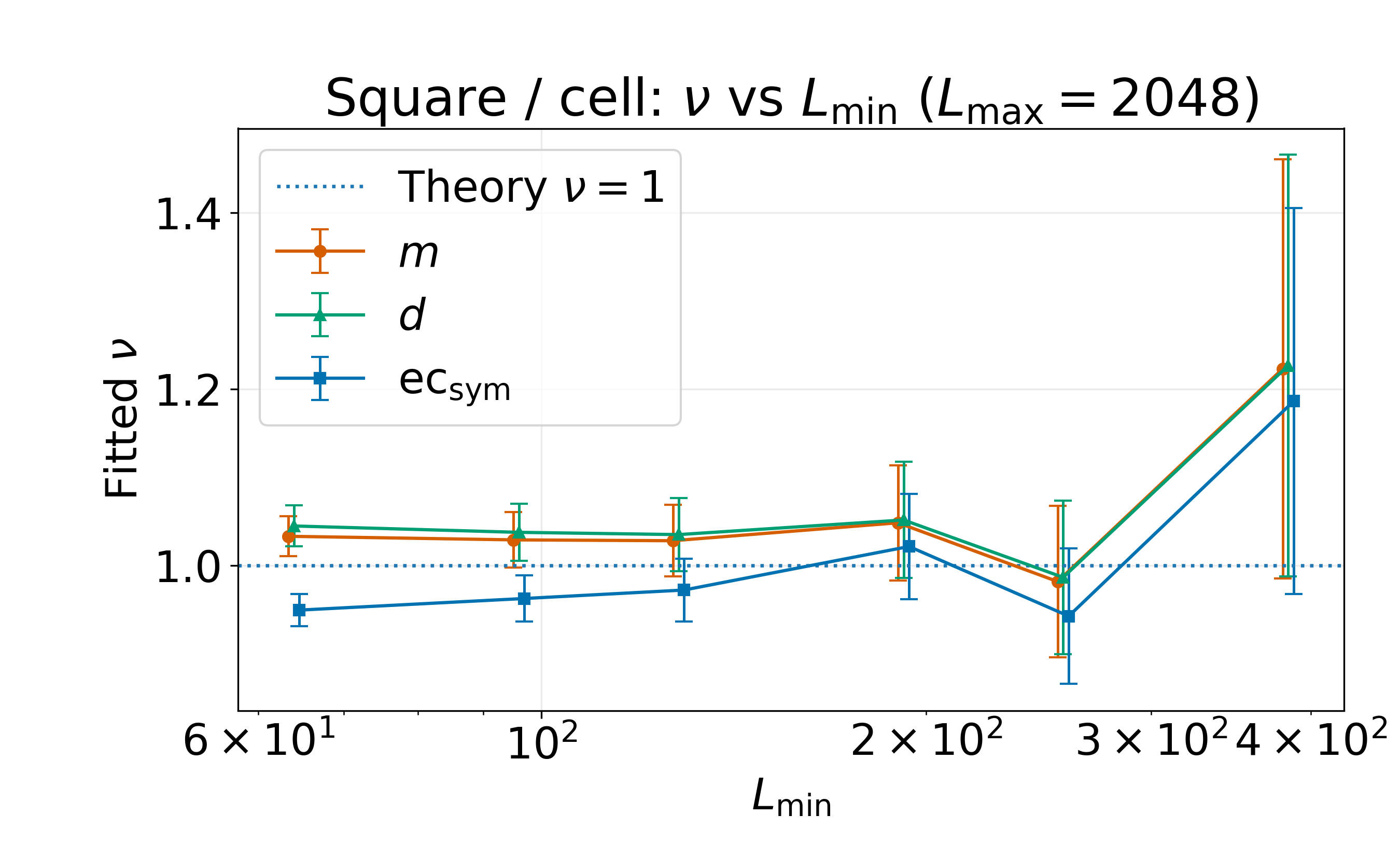}
    \caption{Dependence of the fitted parameters on the lower cutoff $L_{\min}$ used in the finite-size scaling fit for the two-dimensional square lattice. The left and right columns correspond to two choices of the upper cutoff $L_{\max}$, given by the largest and the second-largest available lattice sizes. The top row shows the extrapolated critical temperature $T_c(\infty)$, while the bottom row shows the fitted exponent $\nu$. The horizontal lines indicate the exact two-dimensional Ising values. A small horizontal offset is added to the data points only to improve readability of the error bars.}
    \label{fig:fit_window_stability}
\end{figure*}

Figure~\ref{fig:fit_window_stability} shows the resulting estimates of $T_c(\infty)$ and $\nu$ for the magnetization, $\ecsp$, and $d$. For the critical temperature, the estimates obtained from all three observables remain close to the theoretical value across the considered fitting windows. The observables show small window-dependent variations, but these variations do not change the overall conclusion that both $\ecsp$ and $d$ give consistent estimates of the transition temperature. The estimates obtained from $d$ generally follow those obtained from the magnetization closely.

The differences between the observables are more pronounced for the critical exponent. The estimates obtained from the magnetization and $d$ remain close to each other and to the theoretical value $\nu=1$ over the considered fitting windows, whereas the central values extracted from $\ecsp$ show larger deviations on the available lattice sizes. When smaller lattices are excluded, the fits are based on fewer points and the uncertainty intervals become wider. For fits with larger lower cutoffs, the theoretical value falls within the uncertainty range of the $\ecsp$ estimates, suggesting that the data remain compatible with the expected exponent, although $\ecsp$ does not determine it as sharply as the magnetization and $d$.
\subsection{Extension to the triangular lattice} 
We repeated the same multihistogram reweighting and finite-size scaling analysis for the two-dimensional Ising model on the triangular lattice. This provides a robustness check for the Euler-based construction. The critical temperature depends on the lattice geometry, while the correlation-length exponent is universal within the two-dimensional Ising universality class and is expected to remain $\nu=1$. Table~\ref{tab:fss_tri_Lmin96} summarizes the fit results obtained using the same finite-size scaling form as in Eq.~\eqref{eq:fit_Tc_shift}.

\begin{table}[t]
\centering
\caption{Finite-size scaling results for the 2D triangular lattice. The fitting ranges are selected using the goodness-of-fit criterion $Q\geq 0.05$.}
\label{tab:fss_tri_Lmin96}

\scriptsize
\renewcommand{\arraystretch}{1.15}
\setlength{\tabcolsep}{2.5pt}

\begin{tabular}{l
                S[table-format=3.0]
                S[table-format=2.0]
                S[table-format=1.6(3)]
                S[table-format=1.3(2)]
                S[table-format=1.3]
                S[table-format=1.3]}
\toprule
Observable & {$L_{\min}$} & {$n$} & {$T_c(\infty)$} & {$\nu$} &
{$\chi^2/\mathrm{dof}$} & {$Q$} \\
\midrule
$\ecsp$ & 96 & 11 & \num{3.640924(112)} & \num{0.985(20)} &
\num{1.666} & \num{0.101} \\
$m$ & 64 & 12 & \num{3.640920(100)} & \num{1.006(16)} &
\num{1.866} & \num{0.052} \\
$d$ & 64 & 12 & \num{3.640906(101)} & \num{1.010(16)} &
\num{1.822} & \num{0.059} \\
\bottomrule
\end{tabular}
\end{table}

The goodness-of-fit criterion gives the same lower cutoff, $L_{\min}=64$, for the magnetization and $d$, whereas the selected window for $\ecsp$ starts at $L_{\min}=96$. Thus, the topological observable requires discarding one additional lattice size, indicating that the smallest triangular-lattice systems are less well described by the leading scaling form for $\ecsp$.
All three observables give mutually consistent estimates of the thermodynamic-limit critical temperature. The values obtained from $m$, $\ecsp$, and $d$ agree within their statistical uncertainties and are close to the expected triangular-lattice critical temperature. This shows that both the topological observable $\ecsp$ and the geometric observable $d$ recover the location of the transition.

The estimates of the correlation-length exponent are also compatible with the Ising value $\nu=1$ within one statistical standard deviation. The magnetization gives $\nu=1.006(16)$, $d$ gives $\nu=1.010(16)$ and $\ecsp$ gives $\nu=0.985(20)$. In contrast to the square-lattice results, the triangular-lattice data, therefore show no appreciable deviation of the central value of $\nu_{\ecsp}$ from the expected Ising value. The main distinction is instead the slightly more restrictive fitting window selected for $\ecsp$.

Overall, the triangular-lattice analysis shows that both $\ecsp$ and $d$ provide reliable signatures of the transition. They yield thermodynamic-limit critical temperatures consistent with the magnetization benchmark and exponent estimates compatible with the universal Ising value. For $\ecsp$, the need to discard one additional lattice size indicates that finite-size corrections remain more relevant for the smallest systems, even though the resulting exponent estimate is fully consistent with $\nu=1$.

\section{Discussion}\label{sec:discuss}

A natural question is why, on the square lattice, the symmetrized Euler characteristic gives a less stable finite-size estimate of the critical exponent than the magnetization, while the observable $d$ gives results of comparable quality to $m$. As discussed above, standard renormalization-group arguments imply the same leading asymptotic shift exponent for the three observables. The observed differences should therefore be associated with finite-size corrections rather than different asymptotic critical behavior.

To verify this interpretation more explicitly, we repeat the fits of the pseudocritical temperatures with the Ising value $\nu=1$ fixed and include the first subleading correction,
\begin{equation}
T_c(L)=T_c+\frac{a}{L}+\frac{b}{L^2}.
\label{eq:fss_subleading}
\end{equation}
The $L^{-2}$ term is included as a phenomenological subleading correction, without attributing it to a specific irrelevant scaling operator or correction-to-scaling exponent. The resulting parameters are reported in Table~\ref{tab:fss_subleading}. The leading coefficient $a$ is consistent across all three observables. By contrast, the subleading coefficient is significantly different from zero for $\ecsp$, with $b_{\ecsp}=18.6(4.5)$, while the corresponding coefficients for $m$ and $d$ are consistent with zero.

To determine whether these coefficients differ significantly between observables, we perform a paired bootstrap analysis, preserving the correlations between observables within each lattice size. The resulting differences and their 95\% confidence intervals are reported in Table~\ref{tab:paired_bootstrap_corrections}. The confidence intervals for both $b_{\ecsp}-b_m$ and $b_{\ecsp}-b_d$ exclude zero, indicating statistically significant differences in the fitted subleading coefficients on the square lattice.

\begin{table}[t]
\centering
\caption{Fits of the pseudocritical temperatures on the two-dimensional square lattice to Eq.~\eqref{eq:fss_subleading}, with the Ising value $\nu=1$ fixed and $L_{\min}=64$.}
\label{tab:fss_subleading}

\scriptsize
\renewcommand{\arraystretch}{1.1}
\setlength{\tabcolsep}{4pt}
\begin{tabular}{l
                S[table-format=1.6(2)]
                S[table-format=1.3(2)]
                S[table-format=-2.1(2)]
                S[table-format=1.3]
                S[table-format=1.3]}
\toprule
Observable &
{$T_c(\infty)$} &
{$a$} &
{$b$} &
{$\chi^2/\mathrm{dof}$} &
{$p$} \\
\midrule
$\ecsp$ & \num{2.269136(56)} & \num{2.090(49)} & \num{17.8(44)}  & \num{1.238} & \num{0.266} \\
$m$    & \num{2.269133(56)} & \num{2.090(48)} & \num{-2.8(42)}  & \num{1.067} & \num{0.384} \\
$d$    & \num{2.269132(56)} & \num{2.091(48)} & \num{-5.1(42)}  & \num{1.054} & \num{0.394} \\
\bottomrule
\end{tabular}
\end{table}

\begin{table}[t]
    \centering
    \caption{Differences between the subleading correction coefficients for the square lattice, obtained from paired bootstrap resampling of the fits to Eq.~\eqref{eq:fss_subleading} with $L_{\min}=64$. The intervals correspond to the 95\% bootstrap percentile confidence intervals.}
    \label{tab:paired_bootstrap_corrections}
    \renewcommand{\arraystretch}{1.4}
    \setlength{\tabcolsep}{12pt}
    \begin{tabular}{lcc}
        \hline
        Comparison & $\Delta b$ & 95\% CI \\
        \hline
        $b_{\ecsp}-b_m$ & $20.57$ & $[17.57,\,23.94]$ \\
        $b_{\ecsp}-b_d$ & $22.92$ & $[19.64,\,26.68]$ \\
        \hline
\end{tabular}
\end{table}
The microscopic origin of these corrections remains less clear. One possible contributing factor is the different effective resolution of the observables at finite lattice size. The magnetization directly measures the imbalance between positive and negative spins, whereas $\ecs$ characterizes the topology of the spin domains through the alternating sum of their Betti numbers. Many microscopically distinct configurations can therefore share the same value of $\ecs$, and local spin rearrangements may leave it unchanged. The observable $D$, on the other hand, measures the imbalance between interior points of the positive and negative domains and can distinguish certain local rearrangements that leave $\ecs$ unchanged. Differences in the local geometric contributions entering these observables may therefore affect their effective resolution, particularly for smaller lattice sizes.

To probe this difference, we analyze 200 recorded configurations at $T=2.269$, close to the critical temperature of the square-lattice Ising model. All three observables, $M$, $\ecs$, and $D$, are evaluated on the same configurations. Let $K$ denote the number of distinct values taken by an observable in the sampled ensemble. Since these values generally occur with different frequencies, we also define the entropy-based effective number of sampled values,
\begin{equation}
K_{\mathrm{eff}}=\exp(S_{\mathrm{Sh}}), \qquad
S_{\mathrm{Sh}}=-\sum_{i=1}^{K} p_i\log p_i,
\end{equation}
where $p_i$ is the empirical probability of the $i$th distinct value. By construction, $K_{\mathrm{eff}}\leq K$, with equality when all distinct values occur with equal probability.

\begin{figure}[t]
    \centering
    \includegraphics[width=0.92\linewidth]{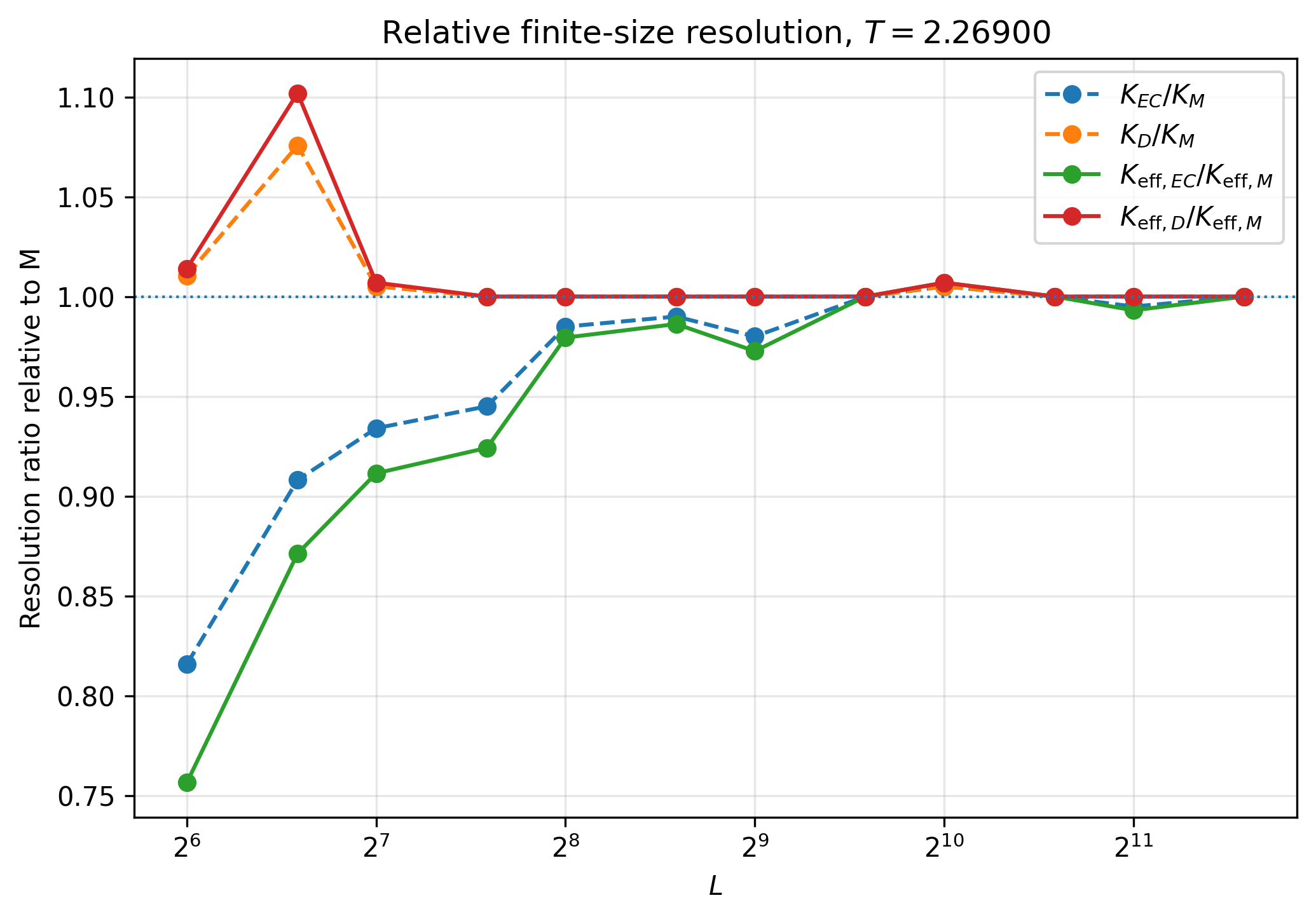}
    \caption{Effective-resolution ratios for the symmetrized Euler characteristic $\ecs$ and the observable $D$, relative to the magnetization. The effective number of sampled values is defined as $K_{\mathrm{eff}}=\exp(S_{\mathrm{Sh}})$, where $S_{\mathrm{Sh}}$ is the Shannon entropy of the empirical distribution. For the smallest lattice sizes, $\ecs$ has a lower effective resolution than $M$, whereas $D$ has a comparable or larger effective resolution. Since only 200 configurations are available at each lattice size, $K_{\mathrm{eff}}\leq 200$; the approach of the ratios toward unity at large $L$ therefore reflects, at least in part, the finite-sample saturation of this measure and should not be interpreted as evidence that the underlying resolutions become equal.}
    \label{fig:distribution_resolution}
\end{figure}

Figure~\ref{fig:distribution_resolution} shows the ratios $K(\ecs)/K(M)$ and $K(D)/K(M)$, together with their entropy-based counterparts $K_{\mathrm{eff}}(\ecs)/K_{\mathrm{eff}}(M)$ and $K_{\mathrm{eff}}(D)/K_{\mathrm{eff}}(M)$. For the smallest lattice sizes, $K_{\mathrm{eff}}(\ecs)/K_{\mathrm{eff}}(M)<1$, showing directly that $\ecs$ distinguishes fewer effective values in the sampled ensemble than the magnetization. In contrast, $D$ has an effective resolution comparable to or larger than that of $M$.

This comparison must, however, be interpreted with an important limitation. Since each empirical distribution is constructed from 200 configurations, $K_{\mathrm{eff}}$ is bounded above by 200 and approaches this value once nearly all sampled configurations yield distinct observable values. Consequently, the apparent convergence of the ratios toward unity at larger $L$ does not demonstrate that the underlying resolutions become equal; rather, the present statistic loses sensitivity once the finite-sample bound is approached. The analysis therefore provides direct evidence for reduced effective resolution of $\ecs$ at small lattice sizes, while remaining inconclusive at larger $L$.

This observation is consistent with the $L_{\min}$ analysis. The largest discrepancy between the estimate of $\nu$ obtained from $\ecs$ and those obtained from $M$ and $D$ occurs when the smallest lattice sizes are included, while the estimates move closer together as these sizes are progressively removed. The explicit subleading fits provide a possible explanation for this behavior: when corrections to the leading finite-size scaling form are neglected, their contribution may be partially absorbed into the fitted exponent, resulting in an effective value of $\nu$ that deviates from its asymptotic value.

The effective-resolution analysis offers an additional perspective on why this effect may be more pronounced for $\ecs$. The smallest lattice sizes are also those for which the differences in effective resolution between $\ecs$ and the other observables are most apparent. This correspondence suggests that the reduced effective resolution of $\ecs$ may contribute to its finite-size behavior, although the present analysis does not establish a causal connection between effective resolution and the subleading corrections. Their microscopic origin therefore remains an open question.

\section{Conclusions}
\label{sec:conclusions}

In this work, we constructed observables from the spin-domain structure of the two-dimensional Ising model by associating cubical and simplicial complexes with the positive and negative spin configurations. From the corresponding Euler characteristics, $\ec^{+}$ and $\ec^{-}$, we introduced the symmetrized Euler characteristic $\ecs$ as a topological observable and considered, in addition, the observable $D$, which captures the imbalance associated with interior points of the spin domains. Both quantities were compared with the standard magnetization through finite-size scaling.

The numerical results show that both $\ecs$ and $D$ display order-parameter-like behavior and provide consistent estimates of the critical temperature. In particular, the pseudocritical temperatures obtained from these observables converge to the expected critical point on both the square and triangular lattices. This shows that both the topological and geometric descriptions of the spin-domain organization individually contain sufficient information to identify the phase transition.

The determination of the critical exponent shows a more differentiated picture. On the square lattice, the estimate of $\nu$ obtained from $\ecs$ exhibits a stronger dependence on the fitting range than the estimates obtained from $M$ and $D$. Fits with $\nu=1$ fixed and a subleading correction
term included are consistent with the data, supporting the interpretation that the larger deviation of the leading-order estimate of $\nu_{\ecs}$ arises from more pronounced finite-size corrections. The distributional analysis provides an additional perspective, revealing a reduced effective resolution of $\ecs$ at smaller lattice sizes, where many microscopically distinct configurations can share the same topological value. This may contribute to the observed finite-size behavior, although the microscopic origin of the corrections remains open. The observable $D$, by contrast, retains an effective number of distinguishable values comparable to or larger than that of the magnetization and yields a more stable estimate of $\nu$. On the triangular lattice, the estimate obtained from $\ecs$ remains consistent with the theoretical Ising value.

An additional theoretical result concerns the averaged Euler characteristic on the square lattice. Its exact relation to the number of local checkerboard configurations allows us to establish a strictly positive lower bound on the mean Euler characteristic density at any nonzero temperature. This rules out the conjecture proposed by Blanchard et al.~\cite{BlanchardEtAl2006}, according to which this density vanishes in the thermodynamic limit throughout the low-temperature phase.

It is important to emphasize that neither $\ecs$ nor $D$ is introduced here as a new independent order parameter for the Ising model. Both observables are strongly related to the magnetic ordering and should instead be viewed as alternative descriptions of the same underlying transition. Their interest lies in the way they separate different aspects of the spin-domain structure: $\ecs$ encodes topological information, while $D$ is sensitive to geometric differences associated with the interiors of the domains. The comparison between them, therefore, provides a way to distinguish which features of the microscopic configuration are retained or compressed by the Euler-characteristic description.

The magnetization should thus be regarded as a benchmark rather than as a competing observable. The Ising model provides a controlled setting in which these constructions can be tested against well-established critical properties. The results show that topology alone is sufficient to recover the location of the transition, while a combined topological and geometric description retains more of the finite-size information relevant for the extraction of critical exponents. This motivates investigating whether analogous topological and geometric observables can provide useful signatures of phase transitions in other systems, including cases in which conventional order parameters are difficult to identify or do not provide a complete description of the transition.

\begin{acknowledgments}

T-SC was partially supported by EPSRC Grant No.~EP/R018472/1 while employed as a Research Officer at Swansea University.
MM was supported by the Dioscuri program initiated by the Max Planck Society, jointly managed with the National Science Centre, Poland and mutually funded by the Polish Ministry of Science and Higher Education and the German Federal Ministry of Research, Technology and Space.
PD and JG were supported by the project ``Centre of Trustworthy Artificial Intelligence for Life Sciences'' implemented under the International Research Agendas programme of the Foundation for Polish Science co-financed by the European Union under the European Funds for Smart Economy 2021--2027 (FENG).
PD acknowledge OpenAI for providing access to ChatGPT Pro used in this work.
The work of BL was partly supported by STFC Consolidated Grants No.~ST/X00063X/1 and No.~UKRI4238. 
The work of DV was partly supported by STFC Consolidated Grants No.~ST/X000680/1 and No.~UKRI/ST/C004237/1.
This research utilized Queen Mary's Apocrita HPC facility, supported by QMUL Research-IT~\cite{Apocrita}.
\end{acknowledgments}

{\bf Open Access Statement.} For the purpose of open access, the authors have applied a Creative Commons Attribution (CC BY) license to any Author Accepted Manuscript version arising.

{\bf Research Data Access Statement.} The data and analysis code supporting this manuscript,as well as the Monte Carlo simulation code, will be made publicly available on Zenodo upon publication of the article (Refs.~\cite{ZENODO1},\cite{ZENODO2}).

\textbf{AI~Declaration.} OpenAI's ChatGPT (GPT-5.6 and GPT-6) and Anthropic's Claude (Fable 5.1) were used to assist with reviewing the manuscript, checking the consistency of the results, and improving the clarity and language of the text. ChatGPT was also used to assist with developing and debugging simulation and analysis scripts, as described in Sec.~\ref{sec:numerical}. The authors reviewed the AI-generated suggestions and take full responsibility for the accuracy of the results, analysis, and interpretations presented in this work.

\section*{Author contributions}
    PD contributed to the original conception of the study, the topological background, and the initial manuscript. BL co-designed the original investigation, suggesting the target model and the use of ensemble averages of topological quantities as phase indicators. BL also contributed to the design of the simulations and analysis strategy, critically discussed the results and their interpretation, drafted the first version of the introduction, and reviewed and commented on the manuscript. DV developed the original Monte Carlo simulation of the Ising model, contributed to a preexisting multihistogram implementation, and contributed to the statistical physics background and the initial manuscript. T-SC further developed the simulation and observable computation code, including the lookup-table method for calculating the Euler characteristic, and contributed to the formulation of the symmetrized Euler characteristic and its extension beyond square lattices. MM extended and corrected the numerical codes, performed the simulations and numerical analysis, derived the theoretical results, prepared the figures, and wrote and revised the majority of the present manuscript. JG supervised the research and contributed through scientific discussions and manuscript review. All authors reviewed and approved the final manuscript.


\bibliography{reference}

@book{kaczynski2004,
  author = {T. Kaczynski and K. Mischaikow and M. Mrozek},
  title = {Computational Homology},
  publisher = {Springer},
  address = {New York},
  year = {2004}
}

@book{Edelsbrunner,
  author    = {Herbert Edelsbrunner and John L. Harer},
  title     = {Computational Topology: An Introduction},
  publisher = {AMS},
  address   = {Providence},
  year      = {2010},
  isbn      = {978-0-8218-4925-5}
}

@article{wolff1989,
  author = {U. Wolff},
  title = {Collective {M}onte {C}arlo updating for spin systems},
  journal = {Phys. Rev. Lett.},
  volume = {62},
  number = {4},
  pages = {361--364},
  year = {1989}
}

@article{Onsager1944,
  author  = {Onsager, Lars},
  title   = {Crystal statistics. {I}. {A} two-dimensional model with an
             order-disorder transition},
  journal = {Phys. Rev.},
  volume  = {65},
  pages   = {117--149},
  year    = {1944},
  doi     = {10.1103/PhysRev.65.117}
}

@article{Houtappel1950,
  author  = {Houtappel, R. M. F.},
  title   = {Order-disorder in hexagonal lattices},
  journal = {Physica},
  volume  = {16},
  number  = {5},
  pages   = {425--455},
  year    = {1950},
  doi     = {10.1016/0031-8914(50)90130-3}
}

@article{FisherBarber1972,
  author  = {Fisher, Michael E. and Barber, Michael N.},
  title   = {Scaling theory for finite-size effects in the critical region},
  journal = {Phys. Rev. Lett.},
  volume  = {28},
  pages   = {1516--1519},
  year    = {1972},
  doi     = {10.1103/PhysRevLett.28.1516}
}

@article{PelissettoVicari2002,
  author  = {Pelissetto, Andrea and Vicari, Ettore},
  title   = {Critical phenomena and renormalization-group theory},
  journal = {Phys. Rep.},
  volume  = {368},
  number  = {6},
  pages   = {549--727},
  year    = {2002},
  doi     = {10.1016/S0370-1573(02)00219-3}
}

@article{KosterlitzThouless1973,
  author  = {Kosterlitz, J. M. and Thouless, D. J.},
  title   = {Ordering, metastability and phase transitions in
             two-dimensional systems},
  journal = {J. Phys. C},
  volume  = {6},
  number  = {7},
  pages   = {1181--1203},
  year    = {1973},
  doi     = {10.1088/0022-3719/6/7/010}
}

@article{FranzosiPettini2004,
  author  = {Franzosi, Roberto and Pettini, Marco},
  title   = {Theorem on the origin of phase transitions},
  journal = {Phys. Rev. Lett.},
  volume  = {92},
  pages   = {060601},
  year    = {2004},
  doi     = {10.1103/PhysRevLett.92.060601}
}

@article{Kastner2008,
  author  = {Kastner, Michael},
  title   = {Phase transitions and configuration space topology},
  journal = {Rev. Mod. Phys.},
  volume  = {80},
  pages   = {167--187},
  year    = {2008},
  doi     = {10.1103/RevModPhys.80.167}
}

@article{CimasoniDelabays2019,
  author  = {Cimasoni, David and Delabays, Robin},
  title   = {The topological hypothesis for discrete spin models},
  journal = {J. Stat. Mech.: Theory Exp.},
  volume  = {2019},
  number  = {3},
  pages   = {033216},
  year    = {2019},
  doi     = {10.1088/1742-5468/ab0c14}
}

@article{Donato2016,
  author  = {Donato, Irene and Gori, Matteo and Pettini, Marco
             and Petri, Giovanni and De Nigris, Sarah and
             Franzosi, Roberto and Vaccarino, Francesco},
  title   = {Persistent homology analysis of phase transitions},
  journal = {Phys. Rev. E},
  volume  = {93},
  pages   = {052138},
  year    = {2016},
  doi     = {10.1103/PhysRevE.93.052138},
  eprint  = {1601.03641},
  archivePrefix = {arXiv},
  primaryClass = {cond-mat.stat-mech}
}

@article{Sale2022,
  author  = {Sale, Nicholas and Giansiracusa, Jeffrey and Lucini, Biagio},
  title   = {Quantitative analysis of phase transitions in
             two-dimensional {XY} models using persistent homology},
  journal = {Phys. Rev. E},
  volume  = {105},
  pages   = {024121},
  year    = {2022},
  doi     = {10.1103/PhysRevE.105.024121},
  eprint  = {2109.10960},
  archivePrefix = {arXiv},
  primaryClass = {cond-mat.stat-mech}
}

@article{BlanchardFortunatoGandolfo2002,
  author  = {Blanchard, Philippe and Fortunato, Santo and Gandolfo, Daniel},
  title   = {Euler--{P}oincar\'e characteristic and phase transition
             in the {P}otts model on $\mathbb{Z}^2$},
  journal = {Nucl. Phys. B},
  volume  = {644},
  number  = {3},
  pages   = {495--508},
  year    = {2002},
  doi     = {10.1016/S0550-3213(02)00681-8}
}

@article{BlanchardEtAl2006,
  author  = {Blanchard, Philippe and Dobrovolny, Christophe
             and Gandolfo, Daniel and Ruiz, Jean},
  title   = {On the mean {E}uler characteristic and mean {B}etti numbers
             of the {I}sing model with arbitrary spin},
  journal = {J. Stat. Mech.: Theory Exp.},
  volume  = {2006},
  number  = {3},
  pages   = {P03011},
  year    = {2006},
  doi     = {10.1088/1742-5468/2006/03/P03011}
}

@article{Tran:2021wyg,
    author = "Tran, Quoc Hoan and Chen, Mark and Hasegawa, Yoshihiko",
    title = "{Topological persistence machine of phase transitions}",
    eprint = "2004.03169",
    archivePrefix = "arXiv",
    primaryClass = "cond-mat.stat-mech",
    doi = "10.1103/PhysRevE.103.052127",
    journal = "Phys. Rev. E",
    volume = "103",
    pages = "052127",
    year = "2021"
}

@article{Kashiwa:2021ctc,
    author = "Kashiwa, Kouji and Hirakida, Takehiro and Kouno, Hiroaki",
    title = "{Persistent homology analysis for dense {QCD} effective model with heavy quarks}",
    eprint = "2103.12554",
    archivePrefix = "arXiv",
    primaryClass = "hep-lat",
    doi = "10.3390/sym14091783",
    journal = "Symmetry",
    volume = "14",
    number = "9",
    pages = "1783",
    year = "2022"
}

@article{Olsthoorn:2020xzs,
    author = "Olsthoorn, Bart and Hellsvik, Johan and Balatsky, Alexander V.",
    title = "{Finding hidden order in spin models with persistent homology}",
    eprint = "2009.05141",
    archivePrefix = "arXiv",
    primaryClass = "cond-mat.stat-mech",
    doi = "10.1103/physrevresearch.2.043308",
    journal = "Phys. Rev. Res.",
    volume = "2",
    number = "4",
    pages = "043308",
    year = "2020"
}

@article{Cole:2020hjx,
    author = "Cole, Alex and Loges, Gregory J. and Shiu, Gary",
    title = "{Quantitative and interpretable order parameters for phase transitions from persistent homology}",
    eprint = "2009.14231",
    archivePrefix = "arXiv",
    primaryClass = "cond-mat.stat-mech",
    doi = "10.1103/PhysRevB.104.104426",
    journal = "Phys. Rev. B",
    volume = "104",
    number = "10",
    pages = "104426",
    year = "2021"
}

@article{Sehayek:2022lxf,
    author = "Sehayek, Dan and Melko, Roger G.",
    title = "{Persistent homology of $\mathbb{Z}^2$ gauge theories}",
    eprint = "2201.09856",
    archivePrefix = "arXiv",
    primaryClass = "cond-mat.stat-mech",
    doi = "10.1103/PhysRevB.106.085111",
    journal = "Phys. Rev. B",
    volume = "106",
    number = "8",
    pages = "085111",
    year = "2022"
}

@article{Sale:2022qfn,
    author = "Sale, Nicholas and Lucini, Biagio and Giansiracusa, Jeffrey",
    title = "{Probing center vortices and deconfinement in {SU(2)} lattice gauge theory with persistent homology}",
    eprint = "2207.13392",
    archivePrefix = "arXiv",
    primaryClass = "hep-lat",
    doi = "10.1103/PhysRevD.107.034501",
    journal = "Phys. Rev. D",
    volume = "107",
    number = "3",
    pages = "034501",
    year = "2023"
}

@article{Spitz:2022tul,
    author = "Spitz, Daniel and Urban, Julian M. and Pawlowski, Jan M.",
    title = "{Confinement in non-{A}belian lattice gauge theory via persistent homology}",
    eprint = "2208.03955",
    archivePrefix = "arXiv",
    primaryClass = "hep-lat",
    doi = "10.1103/PhysRevD.107.034506",
    journal = "Phys. Rev. D",
    volume = "107",
    number = "3",
    pages = "034506",
    year = "2023"
}

@article{Spitz:2024bqh,
    author = "Spitz, Daniel and Urban, Julian M. and Pawlowski, Jan M.",
    title = "{Topological data analysis of the deconfinement transition in {SU(3)} lattice gauge theory}",
    eprint = "2412.09112",
    archivePrefix = "arXiv",
    primaryClass = "hep-lat",
    reportNumber = "MIT-CTP/5801",
    doi = "10.1103/k2xs-4y67",
    journal = "Phys. Rev. D",
    volume = "111",
    number = "11",
    pages = "114519",
    year = "2025"
}

@article{Crean:2024nro,
    author = "Crean, Xavier and Giansiracusa, Jeffrey and Lucini, Biagio",
    title = "{Topological data analysis of monopole current networks in {U(1)} lattice gauge theory}",
    eprint = "2403.07739",
    archivePrefix = "arXiv",
    primaryClass = "hep-lat",
    doi = "10.21468/SciPostPhys.17.4.100",
    journal = "SciPost Phys.",
    volume = "17",
    number = "4",
    pages = "100",
    year = "2024"
}

@misc{Crean:2026caz,
    author = {Crean, Xavier and Giansiracusa, Jeffrey and Lucini, Biagio},
    title = {Simplicity of confinement in {SU(3)} {Y}ang--{M}ills theory},
    eprint = {2602.10088},
    archivePrefix = {arXiv},
    primaryClass = {hep-lat},
    year = {2026}
}

@misc{Caselle:1993yu,
    author = {Caselle, M. and Gliozzi, F. and Vinti, S.},
    title = {Condensation of handles in the interface of 3-{D} {I}sing model},
    eprint = {hep-th/9304001},
    archivePrefix = {arXiv},
    reportNumber = {DFTT-12-93},
    year = {1993}
}

@article{Edelsbrunner2002,
  author  = {Edelsbrunner, Herbert and Letscher, David and Zomorodian, Afra},
  title   = {Topological persistence and simplification},
  journal = {Discrete Comput. Geom.},
  volume  = {28},
  pages   = {511--533},
  year    = {2002},
  doi     = {10.1007/s00454-002-2885-2}
}

@article{ZomorodianCarlsson2005,
  author  = {Zomorodian, Afra and Carlsson, Gunnar},
  title   = {Computing persistent homology},
  journal = {Discrete Comput. Geom.},
  volume  = {33},
  pages   = {249--274},
  year    = {2005},
  doi     = {10.1007/s00454-004-1146-y}
}

@article{Carlsson2009,
  author  = {Carlsson, Gunnar},
  title   = {Topology and data},
  journal = {Bull. Am. Math. Soc.},
  volume  = {46},
  pages   = {255--308},
  year    = {2009},
  doi     = {10.1090/S0273-0979-09-01249-X}
}

@article{Otter2017,
  author  = {Otter, Nina and Porter, Mason A. and Tillmann, Ulrike and Grindrod, Peter and Harrington, Heather A.},
  title   = {A roadmap for the computation of persistent homology},
  journal = {Eur. Phys. J. Data Sci.},
  volume  = {6},
  pages   = {17},
  year    = {2017},
  doi     = {10.1140/epjds/s13688-017-0109-5},
  eprint  = {1506.08903},
  archivePrefix = {arXiv},
  primaryClass = {math.AT}
}

@article{ChazalMichel2021,
  author  = {Chazal, Fr{\'e}d{\'e}ric and Michel, Bertrand},
  title   = {An introduction to topological data analysis: {F}undamental and practical aspects for data scientists},
  journal = {Front. Artif. Intell.},
  volume  = {4},
  pages   = {667963},
  year    = {2021},
  doi     = {10.3389/frai.2021.667963},
  eprint  = {1710.04019},
  archivePrefix = {arXiv},
  primaryClass = {math.ST}
}

@misc{Apocrita,
  author    = {King, Thomas and Butcher, Simon and Zalewski, Lukasz},
  title     = {Apocrita - {H}igh {P}erformance {C}omputing {C}luster for {Q}ueen {M}ary {U}niversity of {L}ondon},
  year      = {2017},
  publisher = {Zenodo},
  doi       = {10.5281/zenodo.438045},
  url       = {https://doi.org/10.5281/zenodo.438045}
}

@misc{ZENODO1,
  author    = {US},
  title     = {Data release},
  year      = {2026},
  publisher = {Zenodo},
  doi       = {10.5281/zenodo.24274386},
  url       = {https://doi.org/10.5281/zenodo.24274386}
}

@misc{ZENODO2,
  author    = {US},
  title     = {Code Release},
  year      = {2026},
  publisher = {Zenodo},
  doi       = {10.5281/zenodo.24273816},
  url       = {https://doi.org/10.5281/zenodo.24273816}
}

@book{Hatcher2002,
  author    = {Allen Hatcher},
  title     = {Algebraic Topology},
  publisher = {Cambridge University Press},
  address   = {Cambridge},
  year      = {2002},
  isbn      = {9780521795401}
}

@book{LandauBinder,
  author    = {Landau, David P. and Binder, Kurt},
  title     = {A Guide to Monte Carlo Simulations in Statistical Physics},
  edition   = {4},
  publisher = {Cambridge University Press},
  address   = {Cambridge},
  year      = {2014},
  doi       = {10.1017/CBO9781139696463}
}

@incollection{PolitisRomano1992,
  author    = {Politis, Dimitris N. and Romano, Joseph P.},
  title     = {A Circular Block-Resampling Procedure for Stationary Data},
  booktitle = {Exploring the Limits of Bootstrap},
  editor    = {LePage, Raoul and Billard, Lynne},
  publisher = {Wiley},
  address   = {New York},
  pages     = {263--270},
  year      = {1992}
}

@article{FerrenbergSwendsen1989,
    author = {Ferrenberg, Alan M. and Swendsen, Robert H.},
    title = {Optimized {M}onte {C}arlo data analysis},
    journal = {Phys. Rev. Lett.},
    volume = {63},
    pages = {1195--1198},
    year = {1989},
    doi = {10.1103/PhysRevLett.63.1195}
}

@article{KastnerMehta2011,
  author = {Kastner, Michael and Mehta, Dhagash},
  title = {Phase transitions detached from stationary points of the energy landscape},
  journal = {Phys. Rev. Lett.},
  volume = {107},
  pages = {160602},
  year = {2011},
  doi = {10.1103/PhysRevLett.107.160602}
}

@article{GoriEtAl2022,
  author = {Gori, Matteo and Franzosi, Roberto and Pettini, Giulio and Pettini, Marco},
  title = {Topological theory of phase transitions},
  journal = {J. Phys. A},
  volume = {55},
  pages = {375002},
  year = {2022},
  doi = {10.1088/1751-8121/ac7f09}
}

@article{Madras:1988ei,
    author = "Madras, Neal and Sokal, Alan D.",
    title = "{The pivot algorithm: {A} highly efficient {M}onte {C}arlo method for the self-avoiding walk}",
    doi = "10.1007/BF01022990",
    journal = "J. Stat. Phys.",
    volume = "50",
    pages = "109--186",
    year = "1988"
}

\end{document}